\documentclass{elsart}
\usepackage{graphics}
\usepackage{amssymb}

\newcommand{\countzero}{\setcounter{equation}{0}%
         \setcounter{figure}{0}%
        \setcounter{table}{0}}

\begin{document}
\hfill {\tiny HISKP-TH-06/12, FZJ-IKP-TH-2006-14}

\begin{frontmatter}
\title{Hyperon-nucleon interactions -- a chiral effective field theory approach}
\author[1]{Henk Polinder\corauthref{cor}},
\corauth[cor]{Corresponding author.}
\ead{h.polinder@fz-juelich.de}
\author[1]{Johann Haidenbauer},
\author[1,2]{Ulf-G. Mei\ss ner}
\address[1]{Institut f{\"u}r Kernphysik (Theorie), Forschungszentrum J{\"u}lich,\\ D-52425 J{\"u}lich, Germany}
\address[2]{Helmholtz-Institut f{\"u}r Strahlen- und Kernphysik (Theorie), Universit{\"a}t Bonn, D-53115 Bonn, Germany}

\begin{abstract}
We construct the leading order hyperon-nucleon potential in chiral effective
field theory. We show that a good description of the available data is
possible and discuss briefly further improvements of this scheme.
\end{abstract}
\end{frontmatter}

\section{Introduction}
\countzero                               
\label{chap:1}
The derivation of nuclear interactions from chiral Effective Field Theory
(EFT) has been discussed extensively in the literature since the work of
Weinberg \cite{Wei90,Wei91}. For reviews we refer to
\cite{Bed02,Epelbaum:2005pn}. 
The main advantages of this scheme are the possibilities to derive two- and three-
nucleon forces as well as external current operators in a consistent way and  
to improve calculations systematically by going to higher orders in the power counting.

Recently the nucleon-nucleon ($NN$) interaction has been described
to a high precision using chiral EFT \cite{Epe05} (see also \cite{Entem:2003ft}).
In this reference, the power counting is applied to the $NN$ potential, as
originally proposed in \cite{Wei90,Wei91}. The $NN$ potential consists of
pion-exchanges and a series of contact interactions with an increasing number
of derivatives to parameterize the shorter ranged part of the $NN$ force. The
pion-exchanges are treated nonperturbatively. A regularized Lippmann-Schwinger
equation is solved to calculate observable quantities. Note that in contrast
to the original Weinberg scheme, the effective potential is made explicitely
energy-independent as it is important for applications in few-nucleon
systems (for details, see \cite{Epe98}).

The hyperon-nucleon ($YN$) interaction has not been investigated using EFT as
extensively as the $NN$ interaction. Hyperon and nucleon mass shifts in
nuclear matter, using chiral perturbation theory, have been studied in
\cite{Sav96}. These authors used a chiral interaction containing four-baryon contact
terms and pseudoscalar-meson exchanges. Recently, the hypertriton and $\Lambda
d$ scattering were investigated in the framework of an EFT with contact 
interactions \cite{Ham02}. Korpa et al. \cite{Kor01} performed a
next-to-leading order (NLO) EFT analysis of
$YN$ scattering and hyperon mass shifts in nuclear matter. Their tree-level
amplitude contains four-baryon contact terms; pseudoscalar-meson exchanges
were not considered explicitly, but ${\rm SU(3)}_f$ breaking by meson masses
was modeled by incorporating dimension two terms coming from one-pion
exchange. The full scattering amplitude was calculated using the
Kaplan-Savage-Wise resummation scheme \cite{Kap98}. The hyperon-nucleon scattering
data were described successfully for laboratory momenta below 200 MeV, using
12 free parameters. Some aspects of strong $\Lambda N$ scattering in effective
field theory and its relation to various formulations of lattice QCD are
discussed in  \cite{Beane:2003yx}.

In this work we apply the scheme used in \cite{Epe05} to the $YN$
interaction. Analogous to the $NN$ potential, at leading order in the power
counting, the $YN$ potential consists of pseudoscalar-meson (Goldstone boson)
exchanges and four-baryon contact terms, related via ${\rm SU(3)}_f$
symmetry. We solve a regularized coupled channels Lippmann-Schwinger equation
for the leading-order (LO) $YN$ potential, including nonderivative contact
terms and one-pseudoscalar-meson exchange, and fit to the low-energy cross
sections, which are dominated by $S$-waves. Contrary to the $NN$ case, it is not
possible to fit to partial waves, since they can not be extracted from the
incomplete and low-precision $YN$ scattering data. 
We remark that our approach is quite different from \cite{Kor01}.

The contents of this paper are as follows. The effective potential is
developed
in Section~\ref{chap:2}. In Section~\ref{chap:2.1}, we first
give a brief recollection of the underlying power counting for the effective potential.
We then investigate the ${\rm SU(3)}_f$ structure of the four-baryon contact
interactions in 
leading order. This is done in Section \ref{chap:2.2}. Here the lowest order ${\rm SU(3)}_f$-invariant four-baryon contact interactions are given and the corresponding potentials are derived. 
Similar to pion-exchanges in the $NN$ case, the $YN$ potential contains the exchanges of pseudoscalar mesons in general. The lowest order ${\rm SU(3)}_f$-invariant interactions are given in Section \ref{chap:2.3}. Here also the one pseudoscalar meson-exchange potential is derived. 
The coupled channels Lippmann-Schwinger equation is solved for the partial-wave projected potential. 
This integral equation is solved in the LSJ basis. The Lippmann-Schwinger equation and the calculation of observable quantities are discussed in Section  \ref{chap:7}.
Results of the fit to the low-energy $YN$ cross sections are presented in Section \ref{chap:6}. Here we show the empirical and calculated total cross sections, differential cross sections and give the values for the scattering lengths. 
Also, predictions for some $YN$ phase shifts are shown and results for the hypertriton binding energy are presented. 
Finally, the summary presents an overview of the research in this work and an outlook for future investigations. Some technical details, of especially the partial wave projection, the LSJ-matrix elements and their derivations, are given in the appendices.

\section{The effective potential}
\countzero
\label{chap:2}

In this section, we construct in some detail the effective chiral
hyperon-nucleon potential at leading order in the (modified) Weinberg
power counting. This power counting is briefly recalled first. Then,
we construct the minimal set of non-derivative four-baryon interactions
and derive the formulae for the one-Goldstone-boson-exchange contributions.

\subsection{Power counting}
\label{chap:2.1}

In this work, we apply the power counting to the effective hyperon-nucleon
potential $V_{\rm eff}$ which is then injected into a regularized
Lippmann-Schwinger equation to generate the bound and scattering states.
The various terms in the effective potential are ordered according to
\begin{equation}
V_{\rm eff} \equiv V_{\rm eff}(Q,g,\mu) = \sum_\nu Q^\nu \, {\mathcal V}_\nu (Q/\mu ,g)~,
\end{equation}
where $Q$ is the soft scale (either a baryon three-momentum, a
Goldstone boson four-momentum  or a Goldstone boson mass), $g$ is a generic
symbol for the pertinent low--energy constants, $\mu$ a regularization
scale, ${\mathcal V}_\nu$ is a function of order one, and $\nu \ge 0$ is the chiral power. 
It can be expressed as 
\begin{eqnarray}\label{eq:power}
\nu &=& 2 - B + 2L + \sum_i v_i \,\Delta_i ~,\nonumber\\
\Delta_i &=& d_i + {\displaystyle\frac{1}{2}}\, b_i  - 2~,
\end{eqnarray}
with $B$ the number of incoming (outgoing) baryon fields, $L$ counts the
number of Goldstone boson loops, and $v_i$ is the number of vertices with
dimension $\Delta_i$. The vertex dimension is expressed in terms of 
derivatives (or Goldstone boson masses) $d_i$  and the number of internal
baryon fields $b_i$ at the vertex under consideration. The leading order (LO)
potential is given by $\nu = 0$, with $B=2$, $L=0$ and $\Delta_i = 0$. Using
Eq.~(\ref{eq:power}) it is easy to see that this condition is fulfilled
for two types of interactions -- a) non-derivative four-baryon contact terms 
with $b_i = 4$ and $d_i = 0$ and b) one-meson exchange diagrams with the
leading meson-baryon derivative vertices allowed by chiral symmetry ($b_i = 2,
d_i = 1$). At LO, the effective potential is entirely given by these two
types of contributions, which will be discussed in detail in the following chapters.

\subsection{The four-baryon contact terms}
\label{chap:2.2}
The leading order contact term for the nucleon-nucleon ($NN$) interactions is
given by  \cite{Wei90,Epe98}
\begin{eqnarray}
{\mathcal L}&=&C_i\left(\bar{N}\Gamma_i N\right)\left(\bar{N}\Gamma_i N\right)\ ,
\label{eq:2.1}
\end{eqnarray}
where $\Gamma_i$ are the usual elements of the Clifford algebra \cite{Bjo65}
\begin{equation}
\Gamma_1=1 \, , \,\, 
\Gamma_2=\gamma^\mu \, , \,\,   
\Gamma_3=\sigma^{\mu\nu} \, , \,\,  
\Gamma_4=\gamma^\mu\gamma_5  \, , \,\, 
\Gamma_5=\gamma_5 \,\, . 
\label{eq:2.2}
\end{equation}
Considering the large components of the nucleon spinors only, the leading order contact term, Eq. (\ref{eq:2.1}), becomes
\begin{eqnarray}
{\mathcal L}&=&-\left(C_1+C_2\right)\left(\varphi^\dagger_N\varphi_N\right)\left(\varphi^\dagger_N\varphi_N\right)+\left(2C_3+C_4\right)\left(\varphi^\dagger_N \mbox{\boldmath $\sigma$} \varphi_N\right)\left(\varphi^\dagger_N \mbox{\boldmath $\sigma$} \varphi_N\right) \nonumber \\
&\equiv&-\frac{1}{2}C_S\left(\varphi^\dagger_N\varphi_N\right)\left(\varphi^\dagger_N\varphi_N\right)-\frac{1}{2}C_T\left(\varphi^\dagger_N \mbox{\boldmath $\sigma$} \varphi_N\right)\left(\varphi^\dagger_N \mbox{\boldmath $\sigma$} \varphi_N\right)\ ,
\label{eq:2.3}
\end{eqnarray}
where $\varphi_N$ are the large components of the nucleon Dirac spinor and
$C_S$ and $C_T$ are constants that need to be determined by fitting to the experimental data.

\begin{figure}[h]
\begin{center}
\resizebox{13.5cm}{4.5cm}{\includegraphics*[2cm,21.0cm][18cm,26cm]{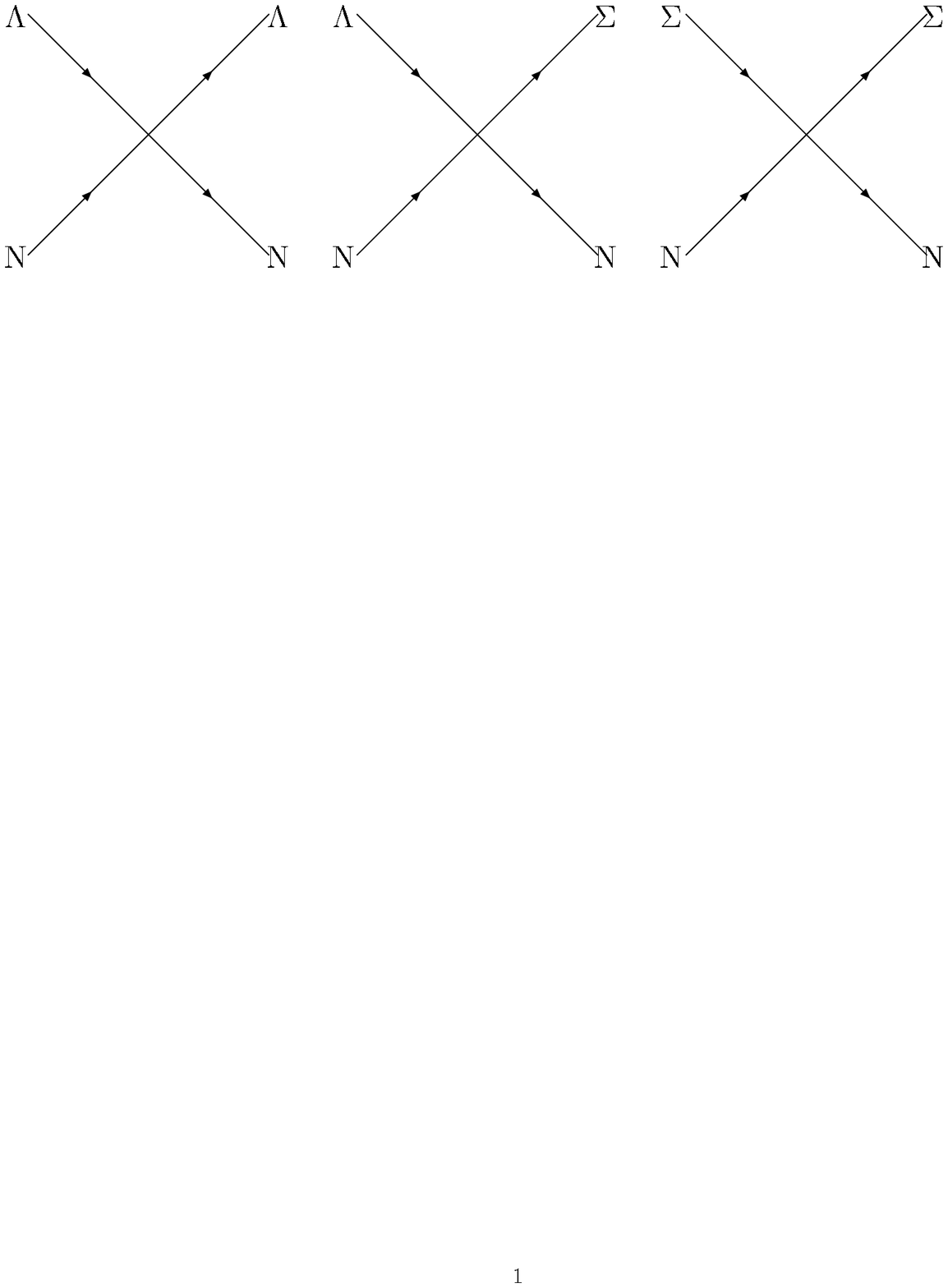}}
\end{center}
\caption{Lowest order contact terms for hyperon-nucleon interactions}
\label{fig:2.0}
\end{figure}

In the case of the hyperon-nucleon ($YN$) interactions we will consider a similar but ${\rm SU(3)}_f$ invariant 
coupling. Thus, let us discuss the flavor structure of the contact terms for the $J^P=\frac{1}{2}^+$ octet baryons 
in the following. The leading order contact terms for the octet baryon-baryon interactions, that are Hermitian 
and invariant under Lorentz transformations, are given by the ${\rm SU(3)}_f$ invariants
\begin{eqnarray}
{\mathcal L}^1&=&{\tilde C}^1_i \left<\bar{B}_a\bar{B}_b\left(\Gamma_i
    B\right)_a\left(\Gamma_i B\right)_b\right>\ , \quad
{\mathcal L}^2 = {\tilde C}^2_i \left<\bar{B}_a\bar{B}_b\left(\Gamma_i B\right)_b\left(\Gamma_i B\right)_a\right>\ ,\nonumber \\
{\mathcal L}^3&=&{\tilde C}^3_i \left<\bar{B}_a\left(\Gamma_i B\right)_a\left(\Gamma_i B\right)_b\bar{B}_b\right>\ ,\quad
{\mathcal L}^4 ={\tilde C}^4_i \left<\bar{B}_a\left(\Gamma_i B\right)_a\bar{B}_b\left(\Gamma_i B\right)_b\right>\ ,\nonumber \\
{\mathcal L}^5&=&{\tilde C}^5_i \left<\bar{B}_a\left(\Gamma_i B\right)_b\bar{B}_b\left(\Gamma_i B\right)_a\right>\ ,\quad
{\mathcal L}^6 = {\tilde C}^6_i \left<\bar{B}_a\left(\Gamma_i B\right)_b\left(\Gamma_i B\right)_a\bar{B}_b\right>\ ,\nonumber \\
&& \nonumber \\
{\mathcal L}^7&=&{\tilde C}^7_i \left<\bar{B}_a\left(\Gamma_i B\right)_a\right>\left<\bar{B}_b\left(\Gamma_i B\right)_b\right>\ ,\quad
{\mathcal L}^8 = {\tilde C}^8_i \left<\bar{B}_a\left(\Gamma_i B\right)_b\right>\left<\bar{B}_b\left(\Gamma_i B\right)_a\right>\ ,\nonumber \\
{\mathcal L}^9&=&{\tilde C}^9_i \left<\bar{B}_a\bar{B}_b\right>\vphantom{\bar{B}_a}\left<\left(\Gamma_i B\right)_a\left(\Gamma_i B\right)_b\vphantom{\bar{B}_a}\right>\ .
\label{eq:2.4}
\end{eqnarray}
Here $a$ and $b$ denote the Dirac indices of the particles, $B$ is the usual irreducible octet representation of ${\rm SU(3)}_f$ given by
\begin{eqnarray}
B&=&
\left(
\begin{array}{ccc}
\frac{\Sigma^0}{\sqrt{2}}+\frac{\Lambda}{\sqrt{6}} & \Sigma^+ & p \\
\Sigma^- & \frac{-\Sigma^0}{\sqrt{2}}+\frac{\Lambda}{\sqrt{6}} & n \\
-\Xi^- & \Xi^0 & -\frac{2\Lambda}{\sqrt{6}}
\end{array}
\right) \ ,
\label{eq:2.5}
\end{eqnarray}
and the brackets $\left< ... \right>$ denote taking the trace in the three-dimensional flavor space. The Clifford algebra elements are here actually diagonal $3\times 3$-matrices in flavor space. Term 9  in Eq. (\ref{eq:2.4}) can be eliminated using the identity
\begin{eqnarray}
&&-\left<\bar{B}_a\bar{B}_b\left(\Gamma_i B\right)_a\left(\Gamma_i B\right)_b\right>+\left<\bar{B}_a\bar{B}_b\left(\Gamma_i B\right)_b\left(\Gamma_i B\right)_a\right> \nonumber \\
&&-\frac{1}{2}\left<\bar{B}_a\left(\Gamma_i B\right)_b\bar{B}_b\left(\Gamma_i B\right)_a\right>+\frac{1}{2}\left<\bar{B}_a\left(\Gamma_i B\right)_a\bar{B}_b\left(\Gamma_i B\right)_b\right> \nonumber \\
&&=\frac{1}{2}\left<\bar{B}_a\left(\Gamma_i B\right)_a\right>\left<\bar{B}_b\left(\Gamma_i B\right)_b\right>-\frac{1}{2}\left<\bar{B}_a\left(\Gamma_i B\right)_b\right>\left<\bar{B}_b\left(\Gamma_i B\right)_a\right> \nonumber \\
&&-\frac{1}{2}\left<\bar{B}_a\bar{B}_b\right>\vphantom{\bar{B}_a}\left<\left(\Gamma_i B\right)_a\left(\Gamma_i B\right)_b\vphantom{\bar{B}_a}\right>\ .
\end{eqnarray}
Making use of the trace property $\left<AB\right>=\left<BA\right>$, we see that the terms 3 and 6 in Eq. (\ref{eq:2.4}) are equivalent to the terms 2 and 1 respectively. Also making use of the Fierz theorem, Appendix \ref{app:A}, one can show that the terms 1, 4 and 8 are equivalent to the terms 2, 5 and 7, respectively. 
So, we only need to consider the terms 2, 5 and 7. Writing these terms explicitly in the isospin basis we find for the $NN$ and $YN$ interactions
\begin{eqnarray}
{\mathcal L}^2&=&{\tilde C}^2_i\left\{ \frac{1}{6}\left[ 5\left(\bar{\Lambda}\Gamma_i \Lambda\right)\left(\bar{N}\Gamma_iN\right)-4\left(\bar{\Lambda}\Gamma_i N\right)\left(\bar{N}\Gamma_i\Lambda\right) \right]  \right. \nonumber \\
&&+\frac{1}{2}\left[ \left(\bar{\mbox{\boldmath $\Sigma$}}\cdot\Gamma_i\mbox{\boldmath $\Sigma$}\right)\left(\bar{N}\Gamma_iN\right) + i\left(\bar{\mbox{\boldmath $\Sigma$}}\times\Gamma_i\mbox{\boldmath $\Sigma$}\right)\cdot\left(\bar{N}\mbox{\boldmath $\tau$}\Gamma_iN\right)  \right] \nonumber \\
&&+ \frac{1}{\sqrt{12}}\left[ \left\{\left(\bar{N}\mbox{\boldmath $\tau$}\Gamma_iN\right)\cdot\left(\bar{\Lambda}\Gamma_i\mbox{\boldmath $\Sigma$}\right)+H.c.\right\} \right. \nonumber \\
&&\left.\left. -2\left\{\left(\bar{N}\Gamma_i\mbox{\boldmath $\Sigma$}\right)\cdot\left(\bar{\Lambda}\mbox{\boldmath $\tau$}\Gamma_iN\right)+H.c.\right\} \right] \vphantom{\frac{1}{6}}\right\} 
\ , \nonumber \\
{\mathcal L}^5&=&{\tilde C}^5_i\left\{\vphantom{\frac{1}{\sqrt{3}}} -\frac{1}{3}\left[ \left(\bar{\Lambda}\Gamma_i \Lambda\right)\left(\bar{N}\Gamma_iN\right)+4\left(\bar{\Lambda}\Gamma_i N\right)\left(\bar{N}\Gamma_i\Lambda\right) \right]  \right. \nonumber \\
&&-\left[ \left(\bar{\mbox{\boldmath $\Sigma$}}\cdot\Gamma_i\mbox{\boldmath $\Sigma$}\right)\left(\bar{N}\Gamma_iN\right) - i\left(\bar{\mbox{\boldmath $\Sigma$}}\times\Gamma_i\mbox{\boldmath $\Sigma$}\right)\cdot\left(\bar{N}\mbox{\boldmath $\tau$}\Gamma_iN\right)  \right] \nonumber \\
&&\left. -\frac{1}{\sqrt{3}}\left[ \left(\bar{N}\mbox{\boldmath $\tau$}\Gamma_iN\right)\cdot\left(\bar{\Lambda}\Gamma_i\mbox{\boldmath $\Sigma$}\right)+H.c. \right] -\left(\bar{N}\Gamma_iN\right)\left(\bar{N}\Gamma_iN\right)\right\} 
\ , \nonumber \\
{\mathcal L}^7&=&{\tilde C}^7_i\left\{
  2\left(\bar{\Lambda}\Gamma_i\Lambda\right)\left(\bar{N}\Lambda
    N\right)+2\left(\bar{\mbox{\boldmath
        $\Sigma$}}\cdot\Gamma_i\mbox{\boldmath
      $\Sigma$}\right)\left(\bar{N}\Gamma_i N\right) 
+\left(\bar{N}\Gamma_iN\right)\left(\bar{N}\Gamma_iN\right)  \right\}\ . \nonumber\\&&
\label{eq:2.6}
\end{eqnarray}
Here $H.c.$ denotes the Hermitian conjugate of the specific term. Also we have introduced the isospinors and isovector according to
\begin{equation}
N=\left(\begin{array}{r}p\\n\end{array}\right)\ ,\ \ \Xi=\left(\begin{array}{r}\Xi^0\\\Xi^-\end{array}\right)\ .
\label{eq:2.7}
\end{equation}
The phases have been chosen according to \cite{Swa63}, such that the inner product of the isovector $\mbox{\boldmath $\Sigma$}$ is
\begin{equation}
\mbox{\boldmath $\Sigma$}\cdot\mbox{\boldmath $\Sigma$}=\Sigma^+\Sigma^-+\Sigma^0\Sigma^0+\Sigma^-\Sigma^+\ .
\label{eq:2.8}
\end{equation}
In order to find the interaction Lagrangian in a more symmetric form (with respect to Fierz rearranged terms like $\left(\bar{\Lambda}\Gamma_i N\right)\left(\bar{N}\Gamma_i\Lambda\right)$) we add to ${\mathcal L}^2$ and ${\mathcal L}^5$ their Fierz rearranged versions and perform a Fierz rearrangement. We find for the $NN$ and $YN$ interactions the Lagrangians
\begin{eqnarray}\label{eq:2.9}
{\mathcal L}^2&=&\frac{C^2_i}{2}\left\{\vphantom{\frac{1}{\sqrt{3}}} \frac{1}{3}\left(\bar{\Lambda}\Gamma_i \Lambda\right)\left(\bar{N}\Gamma_iN\right) +\left[ \left(\bar{\mbox{\boldmath $\Sigma$}}\cdot\Gamma_i\mbox{\boldmath $\Sigma$}\right)\left(\bar{N}\Gamma_iN\right) \right.\right. \nonumber \\
&&\left.\left.+ i\left(\bar{\mbox{\boldmath $\Sigma$}}\times\Gamma_i\mbox{\boldmath $\Sigma$}\right)\cdot\left(\bar{N}\mbox{\boldmath $\tau$}\Gamma_iN\right)  \right] -\frac{1}{\sqrt{3}} \left[\left(\bar{N}\mbox{\boldmath $\tau$}\Gamma_iN\right)\cdot\left(\bar{\Lambda}\Gamma_i\mbox{\boldmath $\Sigma$}\right)+H.c.\right] \right\} \ , \nonumber \\
{\mathcal L}^5&=&-C^5_i\left\{ \frac{5}{3} \left(\bar{\Lambda}\Gamma_i \Lambda\right)\left(\bar{N}\Gamma_iN\right)+\left[ \left(\bar{\mbox{\boldmath $\Sigma$}}\cdot\Gamma_i\mbox{\boldmath $\Sigma$}\right)\left(\bar{N}\Gamma_iN\right) \right.\right. \nonumber \\
&&\left.- i\left(\bar{\mbox{\boldmath $\Sigma$}}\times\Gamma_i\mbox{\boldmath $\Sigma$}\right)\cdot\left(\bar{N}\mbox{\boldmath $\tau$}\Gamma_iN\right)  \right] +\frac{1}{\sqrt{3}}\left[ \left(\bar{N}\mbox{\boldmath $\tau$}\Gamma_iN\right)\cdot\left(\bar{\Lambda}\Gamma_i\mbox{\boldmath $\Sigma$}\right)+H.c. \right] \nonumber \\
&&\left.+\left(\bar{N}\Gamma_iN\right)\left(\bar{N}\Gamma_iN\right)\vphantom{\frac{5}{3}}\right\} \ , \nonumber \\
{\mathcal L}^7&=&C^7_i\left\{
  2\left(\bar{\Lambda}\Gamma_i\Lambda\right)\left(\bar{N}\Lambda
    N\right)+2\left(\bar{\mbox{\boldmath
        $\Sigma$}}\cdot\Gamma_i\mbox{\boldmath
      $\Sigma$}\right)\left(\bar{N}\Gamma_i N\right) 
+\left(\bar{N}\Gamma_iN\right)\left(\bar{N}\Gamma_iN\right)  \right\}\ .\nonumber\\&&
\end{eqnarray}
This is the case for the flavor symmetric interaction (i.e. the ${}^1S_0$ wave). For the flavor antisymmetric interaction (i.e. the ${}^3S_1$ wave) we should have subtracted their Fierz rearranged versions.
The leading order $YN$ contact terms given by these interactions are shown diagrammatically in Figure \ref{fig:2.0}.
If we consider again only the large components of the Dirac spinors in Eq. (\ref{eq:2.9}) then we need, similar to Eq. (\ref{eq:2.3}), six contact constants ($C^2_S$, $C^2_T$, $C^5_S$, $C^5_T$, $C^7_S$ and $C^7_T$,) for the $BB$ interactions.
The (leading order) contact term potential resulting from the interaction Lagrangian Eq. (\ref{eq:2.9}) now becomes
\begin{eqnarray}
V^{(0)}&=&C^{BB}_S+C^{BB}_T\ \mbox{\boldmath $\sigma$}_1\cdot\mbox{\boldmath $\sigma$}_2 \ ,
\label{eq:2.10}
\end{eqnarray}
where the coupling constants $C^{BB}_S$ and $C^{BB}_T$ for the flavor symmetric interaction are defined as
\begin{eqnarray}
C^{NN}_{S,T}&=&-C^5_{S,T}+C^7_{S,T}\ , \nonumber \\
C^{\Lambda \Lambda}_{S,T}&=&\frac{1}{6}C^2_{S,T}-\frac{5}{3}C^5_{S,T}+2C^7_{S,T}\ , \nonumber \\
C^{\Lambda \Sigma}_{S,T}&=&-\frac{1}{\sqrt{3}}\left(\frac{C^2_{S,T}}{2}+C^5_{S,T}\right)\ , \nonumber \\
C^{\Sigma \Sigma}_{S,T}&=&\frac{C^2_{S,T}}{2}-C^5_{S,T}+2C^7_{S,T}\ .
\label{eq:2.11}
\end{eqnarray}
For the flavor antisymmetric interaction the coupling constants $C^{BB}_S$ and $C^{BB}_T$ are defined as
\begin{eqnarray}
C^{NN}_{S,T}&=&C^5_{S,T}+C^7_{S,T}\ , \nonumber \\
C^{\Lambda \Lambda}_{S,T}&=&\frac{3}{2}C^2_{S,T}+C^5_{S,T}+2C^7_{S,T}\ , \nonumber \\
C^{\Lambda \Sigma}_{S,T}&=&-\frac{1}{\sqrt{3}}\left(-\frac{3}{2}C^2_{S,T}+C^5_{S,T}\right)\ , \nonumber \\
C^{\Sigma \Sigma}_{S,T}&=&\frac{C^2_{S,T}}{2}-C^5_{S,T}+2C^7_{S,T}\ .
\label{eq:2.11a}
\end{eqnarray}
However, the coupling constants $C^i_{S,T}$ in Eqs. (\ref{eq:2.11}) and (\ref{eq:2.11a}) still need to be multiplied with the isospin factors given in Table \ref{tab:2.1}.
\begin{table}[t]
\caption{The isospin factors for the various contact terms.}
\label{tab:2.1}
\centering
\begin{tabular}{|r|r|r|r|r|}
\hline
Channel &Isospin &$C^2_{S,T}$ &$C^5_{S,T}$ &$C^7_{S,T}$ \\
\hline
$NN\rightarrow NN$ &0 &- &2 &2 \\
                   &1 &-  &2 &2 \\
\hline
$\Lambda N\rightarrow \Lambda N$ &$\frac{1}{2}$ &1 &1 &1 \\
\hline
$\Lambda N\rightarrow \Sigma N$ &$\frac{1}{2}$ &-$\sqrt{3}$ &-$\sqrt{3}$ &- \\
\hline
$\Sigma N\rightarrow \Sigma N$ &$\frac{1}{2}$ &3 &-1 &1 \\
                               &$\frac{3}{2}$ &0 &2 &1 \\
\hline
\end{tabular}
\end{table}
The $NN$ partial wave potentials now become
\begin{eqnarray}
V^{NN}_{1S0}&=&4\pi\left[-2\left(C^5_S-3C^5_T\right)+2\left(C^7_S-3C^7_T\right)\right]=V^{27} , \nonumber \\
V^{NN}_{3S1}&=&4\pi\left[2\left(C^5_S+C^5_T\right)+2\left(C^7_S+C^7_T\right)\right]=V^{10^*} .
\label{eq:2.13a}
\end{eqnarray}
The $YN$ partial wave potentials become for $\Lambda N \rightarrow \Lambda N$
\begin{eqnarray}
V^{\Lambda\Lambda}_{1S0}&=&4\pi\left[\frac{1}{6}\left(C^2_S-3C^2_T\right)-\frac{5}{3}\left(C^5_S-3C^5_T\right)+2\left(C^7_S-3C^7_T\right)\right] \nonumber \\
&&=\frac{1}{10}\left(9V^{27}+V^{8s}\right) , \nonumber \\
V^{\Lambda\Lambda}_{3S1}&=&4\pi\left[\frac{3}{2}\left(C^2_S+C^2_T\right)+\left(C^5_S+C^5_T\right)+2\left(C^7_S+C^7_T\right)\right] \nonumber \\
&&=\frac{1}{2}\left(V^{8a}+V^{10^*}\right) ,
\label{eq:2.13}
\end{eqnarray}
for isospin-3/2 $\Sigma N\rightarrow \Sigma N$
\begin{eqnarray}
V^{\Sigma\Sigma}_{1S0}&=&4\pi\left[-2\left(C^5_S-3C^5_T\right)+2\left(C^7_S-3C^7_T\right)\right]=V^{27} , \nonumber \\
V^{\Sigma\Sigma}_{3S1}&=&4\pi\left[-2\left(C^5_S+C^5_T\right)+2\left(C^7_S+C^7_T\right)\right]=V^{10} ,
\label{eq:2.14}
\end{eqnarray}
for isospin-1/2 $\Sigma N\rightarrow \Sigma N$
\begin{eqnarray}
\widetilde{V}^{\Sigma\Sigma}_{1S0}&=&4\pi\left[\frac{3}{2}\left(C^2_S-3C^2_T\right)+\left(C^5_S-3C^5_T\right)+2\left(C^7_S-3C^7_T\right)\right] \nonumber \\
&&=\frac{1}{10}\left(V^{27}+9V^{8s}\right) , \nonumber \\
\widetilde{V}^{\Sigma\Sigma}_{3S1}&=&4\pi\left[\frac{3}{2}\left(C^2_S+C^2_T\right)+\left(C^5_S+C^5_T\right)+2\left(C^7_S+C^7_T\right)\right] \nonumber \\
&&=\frac{1}{2}\left(V^{8a}+V^{10^*}\right) ,
\label{eq:2.15}
\end{eqnarray}
and for $\Lambda N\rightarrow \Sigma N$
\begin{eqnarray}
V^{\Lambda\Sigma}_{1S0}&=&4\pi\left[\frac{1}{2}\left(C^2_S-3C^2_T\right)+\left(C^5_S-3C^5_T\right)\right]=\frac{3}{10}\left(-V^{27}+V^{8s}\right) , \nonumber \\
V^{\Lambda\Sigma}_{3S1}&=&4\pi\left[-\frac{3}{2}\left(C^2_S+C^2_T\right)+\left(C^5_S+C^5_T\right)\right]=\frac{1}{2}\left(-V^{8a}+V^{10^*}\right) .
\label{eq:2.16}
\end{eqnarray}
The last part of the previous expressions gives explicitly the ${\rm SU(3)}_f$ representation of the potentials. We note that only 5 of the $\{8\}\times\{8\}=\{27\}+\{10\}+\{10^*\}+\{8\}_s+\{8\}_a+\{1\}$ representations are relevant for $NN$ and $YN$ interactions; equivalently, the six contact terms, $C_S^2$, $C_T^2$, $C_S^5$, $C_T^5$, $C_S^7$, $C_T^7$, enter the $NN$ and $YN$ potentials in only 5 different combinations. 
These 5 contact terms need to be determined by a fit to the experimental data. Since the $NN$ data can not be described with a LO EFT, see \cite{Wei90,Epe00a}, we will not consider the $NN$ interaction explicitly. Therefore, we consider the $YN$ partial wave potentials
\begin{eqnarray}
V^{\Lambda\Lambda}_{1S0}&=&C^{\Lambda\Lambda}_{1S0}\ , \quad
V^{\Lambda\Lambda}_{3S1} = C^{\Lambda\Lambda}_{3S1}\ , \nonumber \\
V^{\Sigma\Sigma}_{1S0}&=&C^{\Sigma\Sigma}_{1S0}\ , \quad
V^{\Sigma\Sigma}_{3S1} = C^{\Sigma\Sigma}_{3S1}\ , \nonumber \\
\widetilde{V}^{\Sigma\Sigma}_{1S0}&=&9C^{\Lambda\Lambda}_{1S0}-8C^{\Sigma\Sigma}_{1S0}\
, \quad
\widetilde{V}^{\Sigma\Sigma}_{3S1} = C^{\Lambda\Lambda}_{3S1}\ , \nonumber \\
V^{\Lambda\Sigma}_{1S0}&=&3\left(C^{\Lambda\Lambda}_{1S0}-C^{\Sigma\Sigma}_{1S0}\right)
\ , \quad
V^{\Lambda\Sigma}_{3S1} = C^{\Lambda\Sigma}_{3S1}\ .
\end{eqnarray}
We have chosen to search for $C^{\Lambda \Lambda}_{1S0}$, $C^{\Lambda \Lambda}_{3S1}$, 
$C^{\Sigma \Sigma}_{1S0}$, $C^{\Sigma \Sigma}_{1S0}$, and $C^{\Lambda \Sigma}_{3S1}$ in the fitting procedure. The other three partial wave potentials are then determined by ${\rm SU(3)}_f$-symmetry.

\subsection{One pseudoscalar-meson exchange}
\label{chap:2.3}
The lowest order ${\rm SU(3)}_f$-invariant pseudoscalar-meson-baryon interaction Lagrangian with the appropriate symmetries is given by (see, e.g., \cite{Mei93}),
\begin{eqnarray}
{\mathcal L}&=&\left< i\bar{B}\gamma^\mu D_\mu B -M_0\bar{B}B+\frac{D}{2}\bar{B}\gamma^\mu\gamma_5 \left\{u_\mu,B\right\} +\frac{F}{2}\bar{B}\gamma^\mu\gamma_5 \left[u_\mu,B\right] \right> \ ,
\label{eq:3.1}
\end{eqnarray}
with $M_0$ the octet baryon mass in the chiral limit. 
There are two possibilities for coupling the axial vector $u_\mu$ to the baryon bilinear. The conventional coupling constants $F$ and $D$, used here, satisfy the relation $F+D=g_A\simeq 1.26$. The axial-vector strength $g_A$ is measured in neutron $\beta$--decay. The covariant derivative acting on the baryons is
\begin{eqnarray}
D_\mu B&=&\partial_\mu B+\left[\Gamma_\mu,B\right] \ ,\nonumber \\
\Gamma_\mu&=&\frac{1}{2}\left[u^\dagger\partial_\mu u+u\partial_\mu u^\dagger\right] \ , \nonumber \\
u^2&=&U=\exp (2iP/\sqrt{2}F_\pi) \ ,
\label{eq:3.2}
\end{eqnarray}
where $F_\pi$ is the weak pion decay constant, $F_\pi =  92.4$ MeV, and $P$ is
the irreducible octet representation of ${\rm SU(3)}_f$ for the pseudoscalar
mesons (the Goldstone bosons)
\begin{eqnarray}
P&=&
\left(
\begin{array}{ccc}
\frac{\pi^0}{\sqrt{2}}+\frac{\eta}{\sqrt{6}} & \pi^+ & K^+ \\
\pi^- & \frac{-\pi^0}{\sqrt{2}}+\frac{\eta}{\sqrt{6}} & K^0 \\
-K^- & \bar{K}^0 & -\frac{2\eta}{\sqrt{6}}
\end{array}
\right) \ .
\label{eq:3.3}
\end{eqnarray}
Symmetry breaking in the decay constants, e.g. $F_\pi \neq F_K$, formally
appears at NLO and will not be considered in the following.
The axial-vector $u_\mu$ is defined as
\begin{eqnarray}
\frac{1}{2}u_\mu&=&\frac{i}{2}\left(u^\dagger\partial_\mu u-u\partial_\mu u^\dagger\right)=\frac{i}{2}\left\{u^\dagger,\partial_\mu u\right\}=\frac{i}{2}u^\dagger \partial_\mu Uu^\dagger \ .
\label{eq:3.4}
\end{eqnarray}
We remark that the first term in the interaction Lagrangian Eq.~(\ref{eq:3.1}) leads to the Weinberg-Tomozawa terms, while the two last terms will lead to one-pseudoscalar-meson exchanges, which are of interest for the leading order potential. 
To evaluate one-pseudoscalar-meson exchange, we write down Eq.~(\ref{eq:3.4}) explicitly and find for the term with the minimal number of pseudoscalar mesons
\begin{eqnarray}
\frac{1}{2}u_\mu&=&-\frac{\partial_\mu P}{\sqrt{2}F_\pi} \ .
\label{eq:3.5}
\end{eqnarray}
Now we find for the last two terms in Eq. (\ref{eq:3.1}) the derivative coupling interaction Lagrangian, leading to one-pseudoscalar-meson-exchange diagrams,
\begin{eqnarray}
{\mathcal L}&=&\left<\frac{D}{2}\bar{B}\gamma^\mu\gamma_5 \left\{u_\mu,B\right\} +\frac{F}{2}\bar{B}\gamma^\mu\gamma_5 \left[u_\mu,B\right] \right>  \nonumber \\
&=&-\left<\frac{g_A(1-\alpha)}{\sqrt{2}F_\pi}\bar{B}\gamma^\mu\gamma_5 \left\{\partial_\mu P,B\right\} +\frac{g_A\alpha}{\sqrt{2}F_\pi}\bar{B}\gamma^\mu\gamma_5 \left[\partial_\mu P,B\right] \right> \ .
\label{eq:3.6}
\end{eqnarray}
Here we have defined $\alpha=F/(F+D)$ and $g_A=F+D$. Writing this interaction Lagrangian explicitly in the isospin basis, we find
\begin{eqnarray}
{\mathcal L}&=&-f_{NN\pi}\bar{N}\gamma^\mu\gamma_5\mbox{\boldmath $\tau$}N\cdot\partial_\mu\mbox{\boldmath $\pi$} +if_{\Sigma\Sigma\pi}\bar{\mbox{\boldmath $ \Sigma$}}\gamma^\mu\gamma_5\times{\mbox{\boldmath $ \Sigma$}}\cdot\partial_\mu\mbox{\boldmath $\pi$} \nonumber \\
&&-f_{\Lambda\Sigma\pi}\left[\bar{\Lambda}\gamma^\mu\gamma_5{\mbox{\boldmath $ \Sigma$}}+\bar{\mbox{\boldmath $\Sigma$}}\gamma^\mu\gamma_5\Lambda\right]\cdot\partial_\mu\mbox{\boldmath $\pi$}-f_{\Xi\Xi\pi}\bar{\Xi}\gamma^\mu\gamma_5\mbox{\boldmath $\tau$}\Xi\cdot\partial_\mu\mbox{\boldmath $\pi$} \nonumber \\
&&-f_{\Lambda NK}\left[\bar{N}\gamma^\mu\gamma_5\Lambda\partial_\mu K+\bar{\Lambda}\gamma^\mu\gamma_5N\partial_\mu K^\dagger\right]
\nonumber \\&&
-f_{\Xi\Lambda K}\left[\bar{\Xi}\gamma^\mu\gamma_5\Lambda\partial_\mu K_c+\bar{\Lambda}\gamma^\mu\gamma_5\Xi\partial_\mu K_c^\dagger\right]
\nonumber \\&&
-f_{\Sigma NK}\left[\bar{\mbox{\boldmath $ \Sigma$}}\cdot\gamma^\mu\gamma_5\partial_\mu K^\dagger\mbox{\boldmath $\tau$}N+\bar{N}\gamma^\mu\gamma_5\mbox{\boldmath $\tau$}\partial_\mu K\cdot{\mbox{\boldmath $ \Sigma$}}\right]
\nonumber \\&&
-f_{\Sigma \Xi K}\left[\bar{\mbox{\boldmath $ \Sigma$}}\cdot\gamma^\mu\gamma_5\partial_\mu K_c^\dagger\mbox{\boldmath $\tau$}\Xi+\bar{\Xi}\gamma^\mu\gamma_5\mbox{\boldmath $\tau$}\partial_\mu K_c\cdot{\mbox{\boldmath $ \Sigma$}}\right]
-f_{NN\eta_8}\bar{N}\gamma^\mu\gamma_5N\partial_\mu\eta
\nonumber \\&&
-f_{\Lambda\Lambda\eta_8}\bar{\Lambda}\gamma^\mu\gamma_5\Lambda\partial_\mu\eta-f_{\Sigma\Sigma\eta_8}\bar{\mbox{\boldmath $ \Sigma$}}\cdot\gamma^\mu\gamma_5{\mbox{\boldmath $ \Sigma$}}\partial_\mu\eta
-f_{\Xi\Xi\eta_8}\bar{\Xi}\gamma^\mu\gamma_5\Xi\partial_\mu\eta \ .
\label{eq:3.7}
\end{eqnarray}
We have introduced the isospin doublets\\
\begin{equation}
N=\left(\begin{array}{r}p\\n\end{array}\right)\ ,\ \ \Xi=\left(\begin{array}{r}\Xi^0\\\Xi^-\end{array}\right)\ ,\ \ K=\left(\begin{array}{r}K^+\\K^0\end{array}\right)\ ,\ \ K_c=\left(\begin{array}{r}\bar{K}^0\\-K^-\end{array}\right)\ .
\label{eq:3.8}
\end{equation}

\begin{figure}[h]
\begin{center}
\resizebox{13.5cm}{11.78cm}{\includegraphics*[2cm,16.5cm][13cm,26cm]{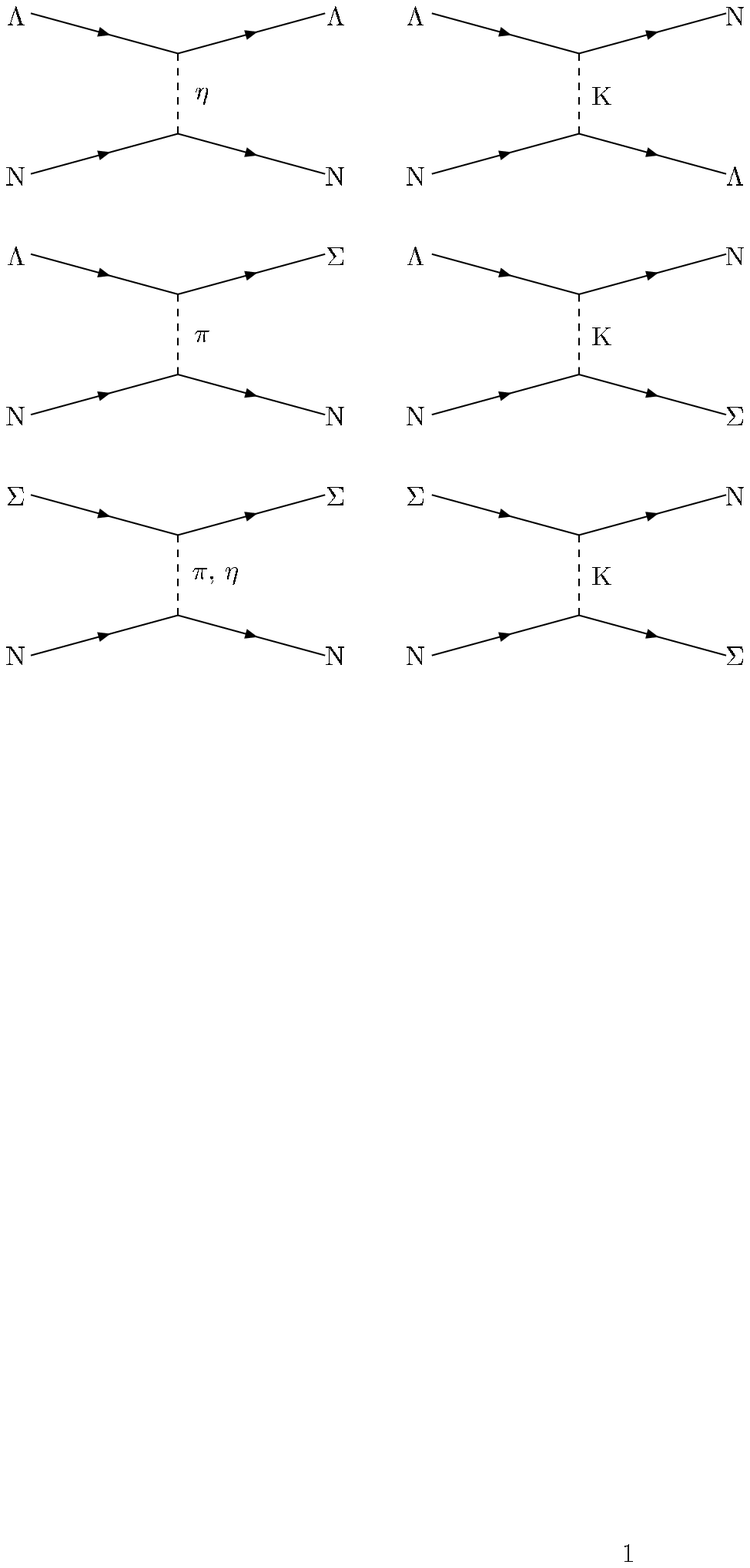}}
\end{center}
\caption{One-pseudoscalar-meson-exchange diagrams for hyperon-nucleon interactions.}
\label{fig:3.0}
\end{figure}

The interaction Lagrangian in Eq. (\ref{eq:3.7}) is invariant under $SU_f(3)$ transformations if the various coupling constants are expressed in terms of the coupling constant $f\equiv g_A/2F_\pi$ and the $F/(F+D)$-ratio $\alpha$ as \cite{Swa63},
\begin{equation}
\begin{array}{rlrlrl}
f_{NN\pi}  = & f, & f_{NN\eta_8}  = & \frac{1}{\sqrt{3}}(4\alpha -1)f, & f_{\Lambda NK} = & -\frac{1}{\sqrt{3}}(1+2\alpha)f, \\
f_{\Xi\Xi\pi}  = & -(1-2\alpha)f, &  f_{\Xi\Xi\eta_8}  = & -\frac{1}{\sqrt{3}}(1+2\alpha )f, & f_{\Xi\Lambda K} = & \frac{1}{\sqrt{3}}(4\alpha-1)f, \\
f_{\Lambda\Sigma\pi}  = & \frac{2}{\sqrt{3}}(1-\alpha)f, & f_{\Sigma\Sigma\eta_8}  = & \frac{2}{\sqrt{3}}(1-\alpha )f, & f_{\Sigma NK} = & (1-2\alpha)f, \\
f_{\Sigma\Sigma\pi}  = & 2\alpha f, &  f_{\Lambda\Lambda\eta_8}  = & -\frac{2}{\sqrt{3}}(1-\alpha )f, & f_{\Xi\Sigma K} = & -f.
\end{array}
\label{eq:3.9}
\end{equation}
Following \cite{Reu94,Hol89,Hai05}, we will neglect the contribution from $\eta$ meson exchange. 
The spin space part of the one-pseudoscalar-meson-exchange potential resulting from the interaction Lagrangian Eq. (\ref{eq:3.7}) is in leading order, similar to the static one-pion-exchange potential (recoil and relativistic corrections give higher order contributions) in \cite{Epe98},
\begin{eqnarray}
V^{(0)}&=&-f_{BBP}^2\frac{\left(\mbox{\boldmath $\sigma$}_1\cdot{\bf k}\right)\left(\mbox{\boldmath $\sigma$}_2\cdot{\bf k}\right)}{{\bf k}^2+\tilde{m}^2}\ ,
\label{eq:3.10}
\end{eqnarray}
where $f_{BBP}$ is one of the coupling constants of Eq. (\ref{eq:3.9}) and $\tilde{m}^2=m^2-\Delta M^2$. Here $m$ is the mass of the exchanged pseudoscalar meson and $\Delta M$ is the baryon mass difference in the energy denominator, which is unequal to zero in the following cases
\begin{eqnarray}
\pi-{\rm exchange,}\ & \Sigma N\rightarrow \Lambda N,\ & \Delta M^2=\left(\frac{M_{\Lambda}-M_{\Sigma}}{2}\right)^2 ,\nonumber \\
K-{\rm exchange,}\ & \Sigma N\rightarrow \Lambda N,\ & \Delta M^2=\left(\frac{M_{\Lambda}+M_{\Sigma}}{2}-M_N\right)^2  ,\nonumber \\
K-{\rm exchange,}\ & \Sigma N\rightarrow \Sigma N,\ & \Delta M^2=\left(M_{\Sigma}-M_N\right)^2  ,\nonumber \\
K-{\rm exchange,}\ & \Lambda N\rightarrow \Lambda N,\ & \Delta M^2=\left(M_{\Lambda}-M_N\right)^2 .
\end{eqnarray} 
Note that these mass shifts are formally of higher order in the chiral
expansion but we include these to have the proper thresholds for the various channels.
Also, we have defined the transferred and average momentum, ${\bf k}$ and ${\bf q}$, in terms of the final and initial center-of-mass (c.m.) momenta of the baryons, ${\bf p}_f$ and ${\bf p}_i$, as
\begin{equation}
{\bf k}={\bf p}_f-{\bf p}_i\ ,\ \ \ {\bf q}=\frac{{\bf p}_f+{\bf p}_i}{2} .
\label{eq:3.11}
\end{equation}
To find the complete (leading order) one-pseudoscalar-meson-exchange potential one needs to multiply the potential in Eq.~(\ref{eq:3.10}) with the isospin factors given in Table~\ref{tab:3.1}.
\begin{table}[t]
\caption{The isospin factors for the various one--pseudoscalar-meson exchanges.}
\label{tab:3.1}
\centering
\begin{tabular}{|r|r|r|r|}
\hline
Channel &Isospin &$\pi$ &$K$ \\
\hline
$NN\rightarrow NN$ &$0$ &$-3$ &$0$ \\
                   &$1$ &$1$  &$0$ \\
\hline
$\Lambda N\rightarrow \Lambda N$ &$\frac{1}{2}$ &$0$ &$1$ \\
\hline
$\Lambda N\rightarrow \Sigma N$ &$\frac{1}{2}$ &$-\sqrt{3}$ &$-\sqrt{3}$ \\
\hline
$\Sigma N\rightarrow \Sigma N$ &$\frac{1}{2}$ &$-2$ &$-1$ \\
                               &$\frac{3}{2}$ &$1$ &$2$ \\
\hline
\end{tabular}
\end{table}
The one-pseudoscalar-meson-exchange diagrams are shown in Figure \ref{fig:3.0}.

\section{Scattering equation and observables}
\countzero
\label{chap:7}
In this section, we briefly comment on the used scattering equation and
the evaluation of observables. The calculations are done in momentum space, 
the scattering equation we solve is the (nonrelativistic) Lippmann-Schwinger
equation. For completeness we briefly discuss it here. The coupled channels 
partial wave Lippmann-Schwinger equation is
\begin{eqnarray}
T_{\rho'\rho}^{\nu'\nu,J}(p',p)&=&V_{\rho'\rho}^{\nu'\nu,J}(p',p) \nonumber\\
&&+ \sum_{\rho'',\nu''}\int_0^\infty \frac{dp''p''^2}{(2\pi)^3} \, V_{\rho'\rho''}^{\nu'\nu'',J}(p',p'') \frac{2\mu_{\nu''}}{p^2-p''^2+i\eta}T_{\rho''\rho}^{\nu''\nu,J}(p'',p)\ .
\nonumber
\end{eqnarray}
The label $\nu$ indicates the particle channels and the label $\rho$ indicates the partial wave. Suppressing the particle channels label, the partial wave projected potentials $V_{\rho'\rho}^{J}(p',p)$ are given in Appendix \ref{app:B}.

The Lippmann-Schwinger equation for the $YN$ system is solved in the particle
basis, in order to incorporate the correct physical thresholds and the Coulomb
interaction in the charged channels. Since the calculations are done in
momentum space, the Coulomb interaction is taken into account according to the
method originally introduced by Vincent and Phatak \cite{Vin74} (see also \cite{Walzl:2000cx}). 
We have used relativistic kinematics for relating the laboratory energy $T_{{\rm lab}}$ of the hyperons to the c.m. momentum. Although we solve the Lippmann-Schwinger equation in the particle basis, the strong potential is calculated in the isospin basis. It contains the leading order contact terms and the one-Goldstone-boson exchanges. The potential in the Lippmann-Schwinger equation is cut off with the regulator function $f^\Lambda(p',p)$,
\begin{equation}
f^\Lambda(p',p)=e^{-\left(p'^4+p^4\right)/\Lambda^4}\ ,
\end{equation}
in order to remove high-energy components of the baryon and pseudoscalar meson fields. 
The differential cross section can be calculated using the (LSJ basis) partial
wave amplitudes, for details we refer to \cite{Nag75,Hol89}. 
The total cross sections are found by simply 
integrating the differential cross sections, except for the 
$\Sigma^+ p\to \Sigma^+ p$ and $\Sigma^- p\rightarrow \Sigma^- p$
channels. For those channels 
the experimental total cross sections were obtained via \cite{Eis71}
\begin{eqnarray}\label{eq:sigtot}
\sigma&=&\frac{2}{\cos \theta_{{\rm max}}-\cos \theta_{{\rm min}}}\int_{\cos \theta_{{\rm min}}}^{\cos \theta_{{\rm max}}}\frac{d\sigma(\theta)}{d\cos \theta}d\cos \theta \ ,
\end{eqnarray}
for various values of $\cos \theta_{{\rm min}}$ and $\cos \theta_{{\rm max}}$. Following \cite{Rij99}, we use $\cos \theta_{{\rm min}}=-0.5$ and $\cos \theta_{{\rm max}}=0.5$ in our calculations for the $\Sigma^+ p\rightarrow \Sigma^+ p$ and $\Sigma^- p\rightarrow \Sigma^- p$ cross sections, in order to stay as close as possible 
to the experimental procedure.

\section{Results and discussion}
\begin{figure}[h]
\begin{center}
\resizebox{14.0cm}{16.47cm}{\includegraphics*[2.0cm,6.8cm][19.5cm,27cm]{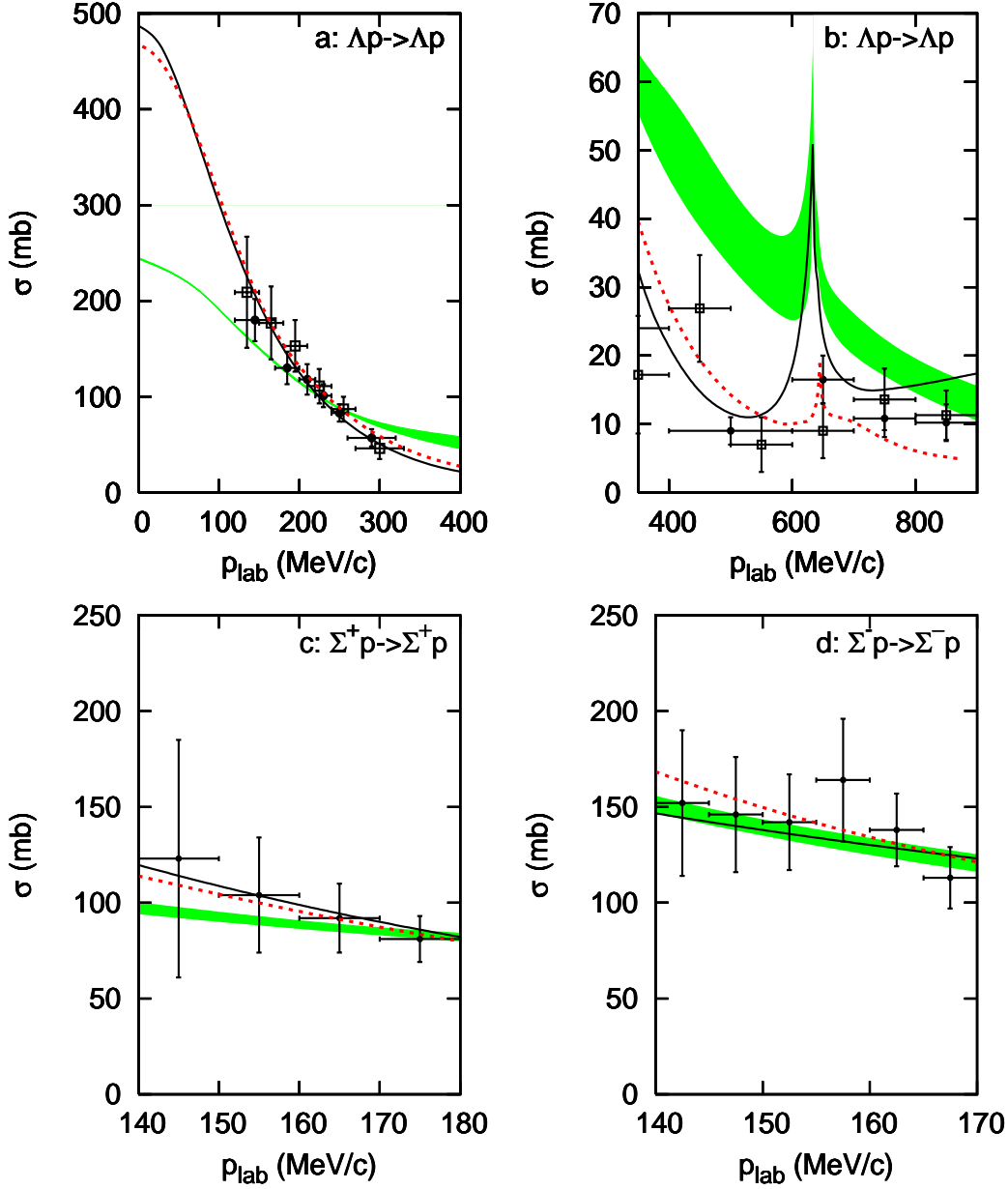}}
\end{center}
\caption{''Total'' cross section $\sigma$ (as defined in Eq.~(\ref{eq:sigtot}))
as a function of $p_{{\rm lab}}$. 
The experimental cross sections in $a$ are taken from Refs.~\cite{Sec68} (open squares) and 
~\cite{Ale68} (filled circles),
in $b$ from Refs.~\cite{Kad71} (filled circles) and \cite{Hau77} (open squares)
and in $c$,$d$ from \cite{Eis71}. 
The shaded band is the J{\"u}lich chiral EFT'06 A for $\Lambda = 550,...,700$ MeV, the dashed curve is the J{\"u}lich '04 model \cite{Hai05}, and the solid curve is the Nijmegen NSC97f model \cite{Rij99}.}
\label{fig:6.0}
\end{figure}

\begin{figure}[h]
\begin{center}
\resizebox{14.0cm}{16.47cm}{\includegraphics*[2cm,6.8cm][19.5cm,27cm]{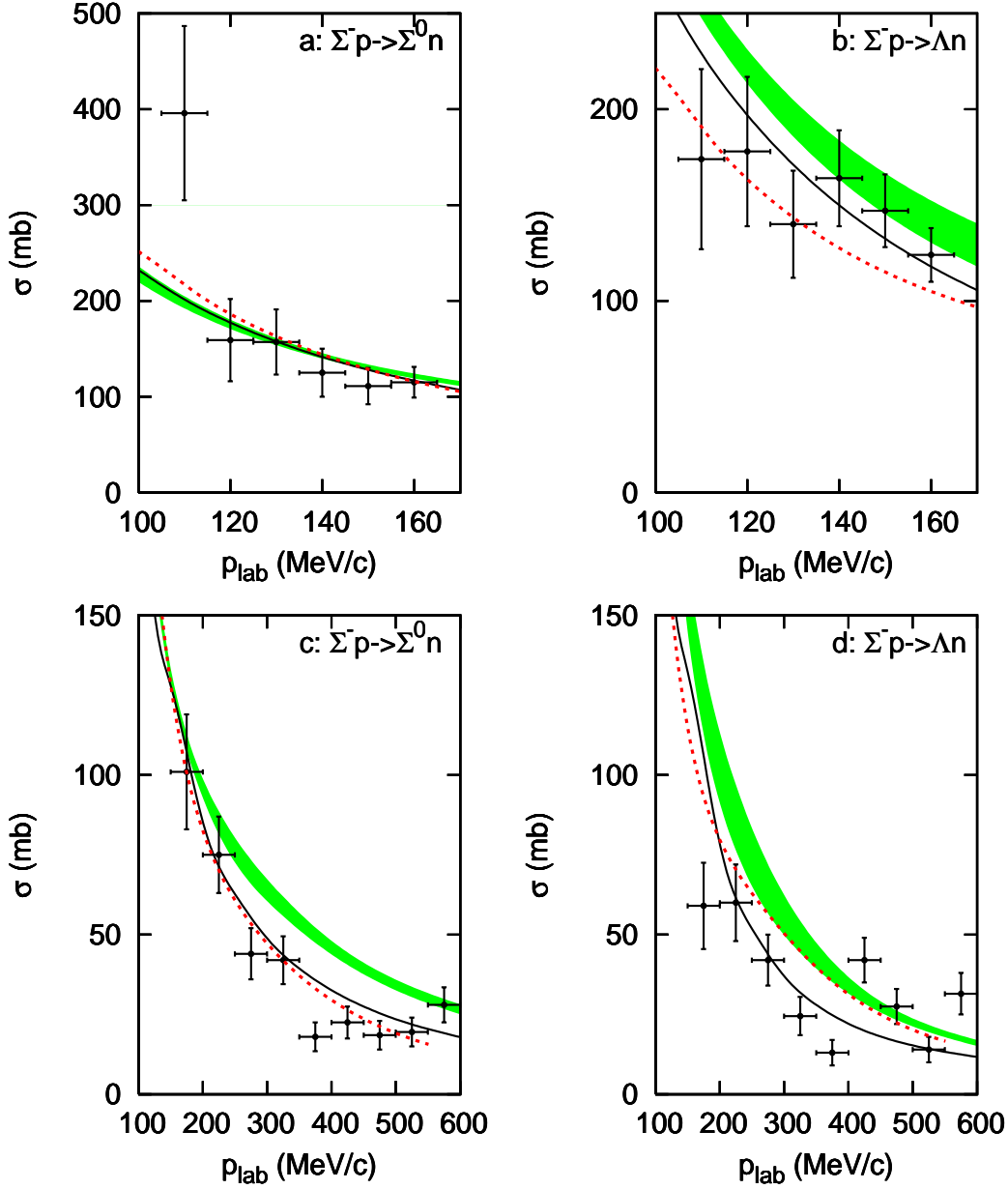}}
\end{center}
\caption{As in Figure \ref{fig:6.0}, but now the experimental cross sections in $a$,$b$
are taken from Refs.~\cite{Eng66} and in $c$,$d$ from \cite{Ste70}.}
\label{fig:6.1}
\end{figure}

\countzero
\label{chap:6}
For the fitting procedure we consider the empirical low-energy total cross sections shown in 
Figures \ref{fig:6.0}a,c, and d and \ref{fig:6.1}a and b, 
and the inelastic capture ratio at rest \cite{Swa62}, in total 35 $YN$ data. 
These data have also been used in \cite{Hai05,Rij99} and are listed in Table \ref{tab:6.1a} (see below). The higher energy total cross sections and differential cross sections are then predictions of the LO chiral EFT, which contains five free parameters. The fits are done for fixed values of the cut--off mass and of $\alpha$, the pseudoscalar $F/(F+D)$ ratio.

The five LECs $C^{\Lambda \Lambda}_{1S0}$, $C^{\Lambda \Lambda}_{3S1}$, $C^{\Sigma \Sigma}_{1S0}$, $C^{\Sigma \Sigma}_{3S1}$, and $ C^{\Lambda \Sigma}_{3S1}$, in Eqs. (\ref{eq:2.13}), (\ref{eq:2.14}), and (\ref{eq:2.16}), were varied during the parameter search to the set of 35 low-energy $YN$ data. The other LECs are then determined by ${\rm SU(3)}_f$ symmetry. The values of the contact terms obtained in the fitting procedure for cut--off values between $550$ and $700$ MeV, are listed in Table \ref{tab:6.1}.
\begin{table}[t]
\caption{The $YN$ $S$-wave contact terms for various cut--offs. The values of 
the LECs are in $10^4$ ${\rm GeV}^{-2}$; the values of $\Lambda$ in MeV. 
$\chi^2$ is the total chi squared for 35 $YN$ data.}
\label{tab:6.1}
\vspace{0.2cm}
\centering
\begin{tabular}{|r|rrrr|}
\hline
$\Lambda$& $550$& $600$& $650$& $700$  \\
\hline
$C^{\Lambda \Lambda}_{1S0}$ &$-.0467$ &$-.0536$ &$-.0520$ &$-.0516$\\
$C^{\Lambda \Lambda}_{3S1}$ &$-.0214$ &$-.0162$ &$-.0097$ &$-.0024$\\
$C^{\Sigma \Sigma}_{1S0}$   &$-.0797$ &$-.0734$ &$-.0738$ &$-.0730$\\
$C^{\Sigma \Sigma}_{3S1}$    &$.0398$  &$.2486$  &$.1232$  &$.1235$\\
$C^{\Lambda \Sigma}_{3S1}$   &$.0035$ &$-.0063$ &$-.0048$ &$-.0025$\\
\hline
\hline
$\chi^2$& 27.8& 29.0& 33.5& 42.8\\
\hline
\end{tabular}
\end{table}
The fits were first done for the cut-off mass $\Lambda =600$ MeV. 
We remark that the $\Lambda N$ $S$-wave scattering lengths resulting for that cut-off were then 
kept fixed in the subsequent fits for the other cut--off values. 
We did this because the $\Lambda N$ scattering lengths are not well determined by the 
scattering data. As a matter of facts, not even the relative magnitude of the $\Lambda N$ triplet 
and singlet interaction can be constrained from the $YN$ data, 
but their strengths play an important role for the hypertriton binding energy \cite{YNN1}. 
Contrary to the $NN$ case, see, e.g. \cite{Epe00a}, the contact terms are in general not determined by a specific phase shift, because of the coupled particle channels in the $YN$ interaction. Furthermore, the limited accuracy and incompleteness of the $YN$ scattering data do not allow for a unique partial wave analysis. Therefore we have fitted the chiral EFT directly to the cross sections.
A comparison between the experimental scattering data considered and the values found in the fitting procedure is given in Table \ref{tab:6.1a}, for $\Lambda = 550$ MeV.
\begin{table}[t]
\caption{Comparison between the 35 experimental $YN$ data and the theoretical
  values for the cut--off $\Lambda = 550$ MeV. Momenta are in units of MeV and 
cross sections in mb.}
\label{tab:6.1a}
\vspace{0.2cm}
\centering
\begin{tabular}{|rrr|rrr|rrr|}
\hline
\multicolumn{3}{|c|}{$\Lambda p \rightarrow \Lambda p\;\;$ $\chi^2 = 7.9$}& \multicolumn{3}{c|}{$\Lambda p \rightarrow \Lambda p\;\;$ $\chi^2 = 4.8$}& \multicolumn{3}{c|}{$\Sigma^- p \rightarrow \Lambda n\;\;$ $\chi^2 = 6.3$}\\
$p_{{\rm lab}}^{\Lambda}$& $\sigma_{{\rm exp}}$\cite{Sec68}& $\sigma_{{\rm the}}$& $p_{{\rm lab}}^{\Lambda}$& $\sigma_{{\rm exp}}$\cite{Ale68}& $\sigma_{{\rm the}}$& $p_{{\rm lab}}^{\Sigma^-}$& $\sigma_{{\rm exp}}$\cite{Eng66}& $\sigma_{{\rm the}}$ \\
\hline
135 &209$\pm$58 &162.8 &145 &180$\pm$22 &154.8 &110 &174$\pm$47 &249.3 \\
165 &177$\pm$38 &139.5 &185 &130$\pm$17 &125.3 &120 &178$\pm$39 &213.8 \\
195 &153$\pm$27 &118.7 &210 &118$\pm$16 &109.5 &130 &140$\pm$28 &185.8 \\
225 &111$\pm$18 &101.0 &230 &101$\pm$12 &98.3  &140 &164$\pm$25 &163.4 \\
255 &87 $\pm$13 &86.1  &250 &83 $\pm$9  &88.4  &150 &147$\pm$19 &145.3 \\
300 &46 $\pm$11 &68.8  &290 &57 $\pm$9  &72.3  &160 &124$\pm$14 &130.4 \\
\hline
\hline
\multicolumn{3}{|c|}{$\Sigma^+ p \rightarrow \Sigma^+ p\;\;$ $\chi^2 = 0.4$}& \multicolumn{3}{c|}{$\Sigma^- p \rightarrow \Sigma^- p\;\;$ $\chi^2 = 1.6$}& \multicolumn{3}{c|}{$\Sigma^- p \rightarrow \Sigma^0 n\;\;$ $\chi^2 = 6.8$}\\
$p_{{\rm lab}}^{\Sigma^-}$& $\sigma_{{\rm exp}}$\cite{Eis71}& $\sigma_{{\rm the}}$& $p_{{\rm lab}}^{\Sigma^-}$& $\sigma_{{\rm exp}}$\cite{Eis71}& $\sigma_{{\rm the}}$& $p_{{\rm lab}}^{\Sigma^-}$& $\sigma_{{\rm exp}}$\cite{Eng66}& $\sigma_{{\rm the}}$ \\
\hline
145 &123$\pm$62 &97.6 &142.5 &152$\pm$38 &152.2 &110 &396$\pm$91 &204.9 \\
155 &104$\pm$30 &92.4 &147.5 &146$\pm$30 &145.6 &120 &159$\pm$43 &180.5 \\
165 &92 $\pm$18 &87.6 &152.5 &142$\pm$25 &139.5 &130 &157$\pm$34 &161.0 \\
175 &81 $\pm$12 &83.0 &157.5 &164$\pm$32 &133.8 &140 &125$\pm$25 &145.3 \\
    &           &     &162.5 &138$\pm$19 &128.5 &150 &111$\pm$19 &132.3 \\
    &           &     &167.5 &113$\pm$16 &123.6 &160 &115$\pm$16 &121.5 \\
\hline
\hline
\multicolumn{3}{|c}{$r^{{\rm exp}}_R = 0.468\pm 0.010$}& \multicolumn{3}{c}{$r^{{\rm the}}_R = 0.465$}& \multicolumn{3}{c|}{$\chi^2 = 0.1$}\\
\hline
\end{tabular}
\end{table}
A good description of the considered $YN$ scattering data has been obtained in the considered cut--off region, as can be seen in Tables \ref{tab:6.1} and \ref{tab:6.1a} and Figures \ref{fig:6.0}a,c,d and \ref{fig:6.1}a,b.
\begin{figure}[h]
\begin{center}
\resizebox{14.0cm}{16.47cm}{\includegraphics*[2cm,6.8cm][19.5cm,27cm]{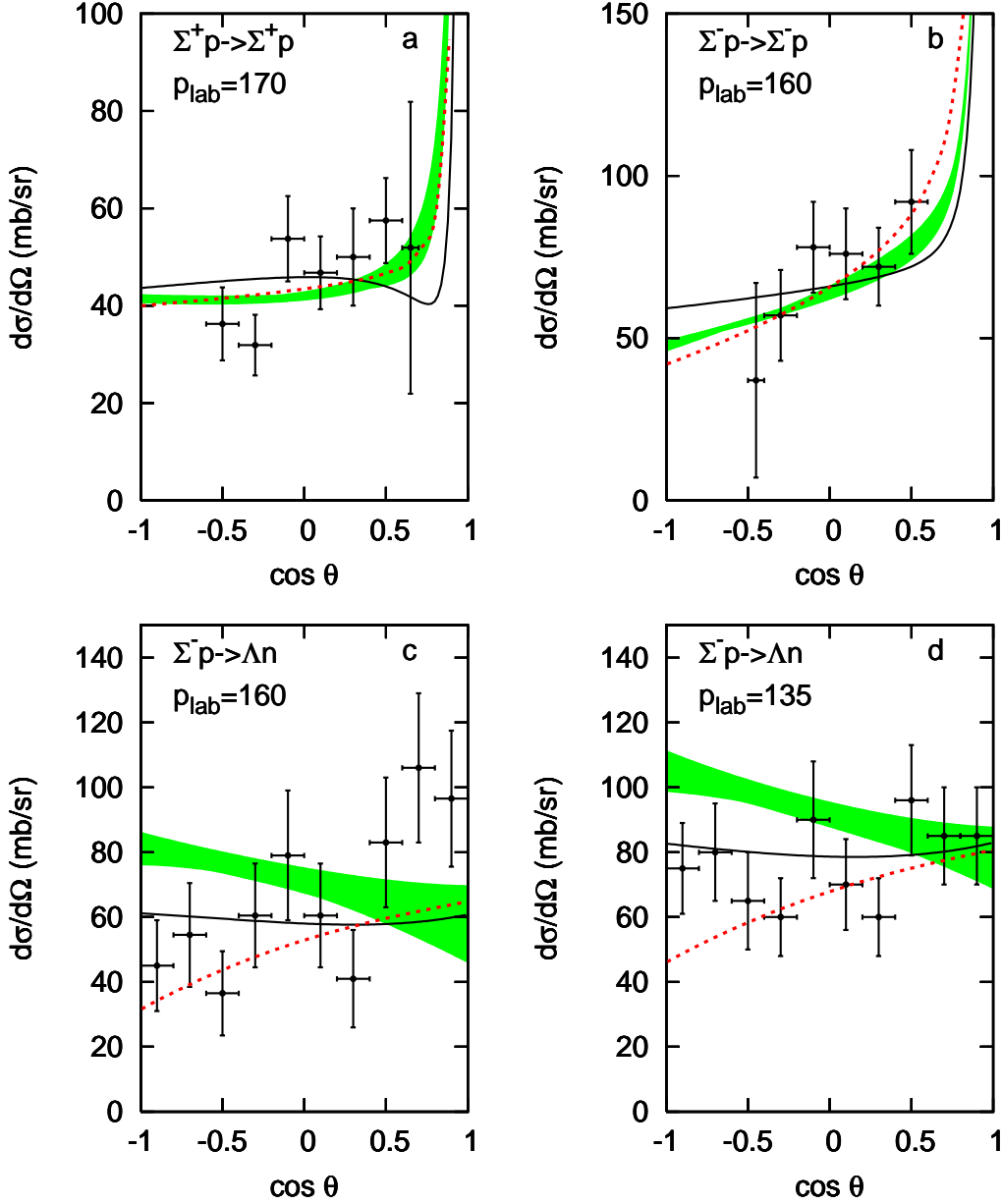}}
\end{center}
\caption{Differential cross section $d\sigma / d\cos \theta$ as a function of $\cos \theta$, where $\theta$ is the c.m. scattering angle, at various values of $p_{{\rm lab}}$ (MeV/c). 
The experimental differential cross sections in $a$,$b$ are taken from \cite{Eis71} and in 
$c$,$d$ from \cite{Eng66}. Same description of curves as in Figure \ref{fig:6.0}.}
\label{fig:6.2}
\end{figure}
\begin{figure}[h]
\begin{center}
\resizebox{14.0cm}{16.47cm}{\includegraphics*[2cm,6.8cm][19.5cm,27cm]{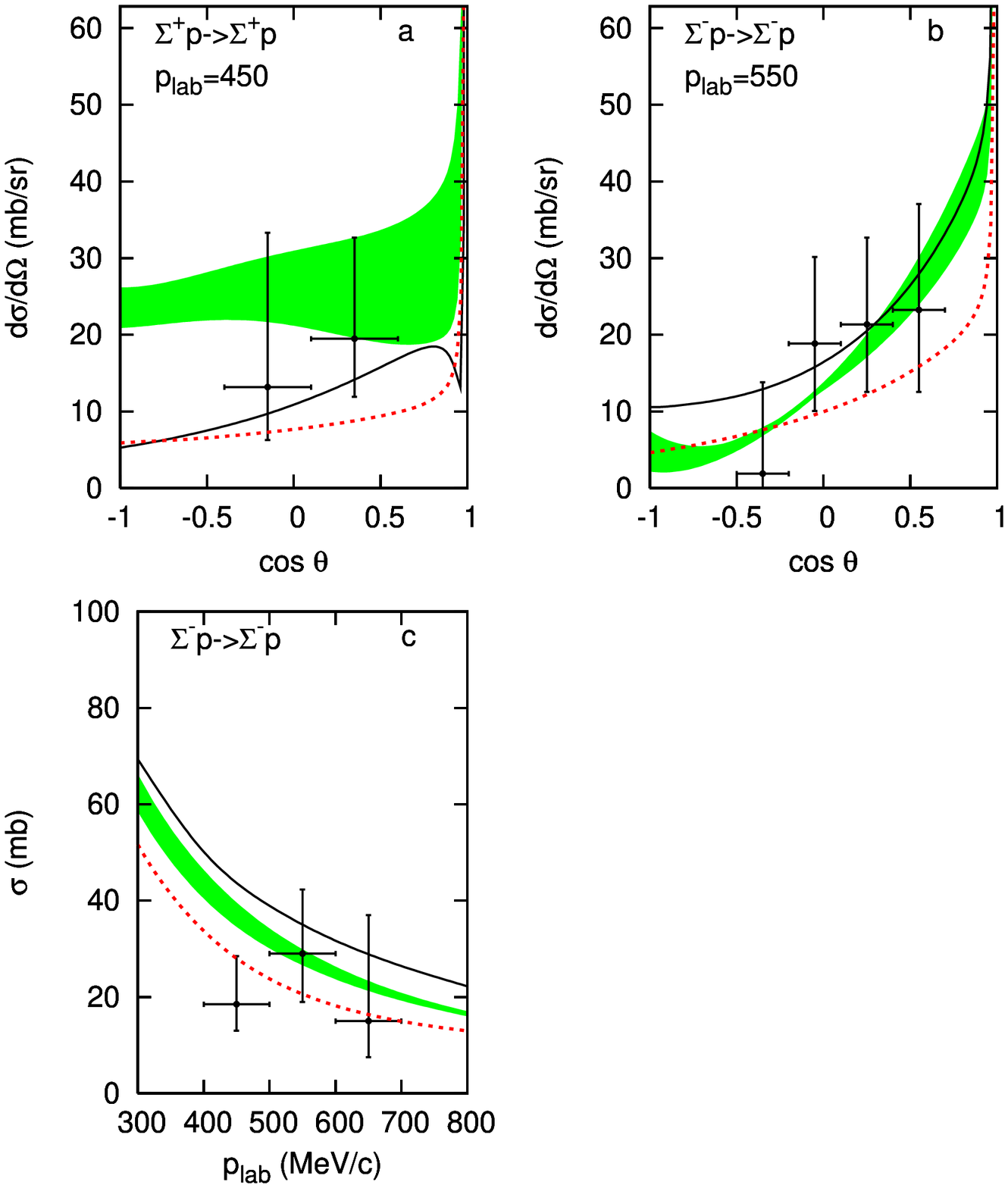}}
\end{center}
\caption{Recent $YN$ data. $a$ and $b$: differential cross section $d\sigma / d\cos \theta$ as a function of $\cos \theta$, where $\theta$ is the c.m. scattering angle, at various values of $p_{{\rm lab}}$ (MeV/c). The experimental differential cross sections are from \cite{Ahn99} and \cite{Kon00}, respectively. 
$c$: ''total'' cross section $\sigma$ as a function of $p_{{\rm lab}}$. The experimental cross sections are 
from \cite{Kon00}. Same description of curves as in Figure \ref{fig:6.0}.}
\label{fig:6.3}
\end{figure}
In these figures the shaded band represents the results of the chiral EFT in the considered cut--off region. 
In this low-energy regime the cross sections are mainly given by the $S$-wave contribution, except for 
for the $\Lambda N\rightarrow \Sigma N$ cross section where 
the ${}^3D_1(\Lambda N)\leftrightarrow {}^3S_1(\Sigma N)$ transition provides the main contribution.
Still all partial waves with total angular momentum $J\le 2$ were included in the computation of the
observables. 
The $\Lambda p$ cross section shows a clear cusp, peaking at 65 mb, at the $\Sigma^+ n$ threshold, see Figure \ref{fig:6.0}b. It is hard to see this effect in the experimental data, since it occurs over a very narrow energy range. Figure \ref{fig:6.0}b shows that the predicted $\Lambda p$ cross section at higher energies is too large, 
which is related to the problem that some LO partial waves are too large at higher energies. Note that 
this was also the case for the $NN$ interaction \cite{Epe00a}. In a NLO calculation this problem will probably vanish. The differential cross sections at low energies, which have not been taken into account in the fitting procedure, are predicted well, see Figure \ref{fig:6.2}. The results of the chiral EFT are also in good agreement with 
the scattering data at higher energy, the older ones in Figures \ref{fig:6.1}c,d as well as the
more recent scattering data in Figure \ref{fig:6.3}.

The $\Lambda p$ and $\Sigma^+p$ scattering lengths and effective ranges are listed in Table \ref{tab:6.2} 
together with the corresponding hypertriton binding energies (preliminary results of $YNN$ Faddeev 
calculations from \cite{Nog06}).
\begin{table}[h]
\caption{The $YN$ singlet and triplet scattering lengths and effective ranges
  (in fm) and the hypertriton binding energy, $E_B$ (in MeV). We notice that the
  deuteron binding energy is $-2.224$~MeV. The binding energies for the 
  hypertriton (last row), \cite{Nog06}, are calculated using 
  the Idaho-N3LO $NN$ potential \cite{Entem:2003ft}. The
  experimental value of the hypertriton binding energy is $-2.354(50)$~MeV.}
\label{tab:6.2}
\vspace{0.2cm}
\centering
\begin{tabular}{|r|rrrr|}
\hline
$\Lambda$& 550& 600& 650& 700  \\
\hline
$a^{\Lambda p}_s$ &$-1.80$ &$-1.80$ &$-1.80$ &$-1.80$ \\
$r^{\Lambda p}_s$  &$1.72$  &$1.76$  &$1.74$  &$1.73$ \\
$a^{\Lambda p}_t$ &$-1.22$ &$-1.23$ &$-1.23$ &$-1.23$ \\
$r^{\Lambda p}_t$  &$2.05$  &$2.16$  &$2.23$  &$2.29$ \\
\hline
$a^{\Sigma^+ p}_s$  &$-2.92$ &$-2.26$  &$-2.48$  &$-2.55$ \\
$r^{\Sigma^+ p}_s$   &$2.80$  &$3.51$   &$3.21$   &$3.12$ \\
$a^{\Sigma^+ p}_t$   &$0.27$  &$0.65$   &$0.49$   &$0.46$ \\
$r^{\Sigma^+ p}_t$ &$-20.37$ &$-2.46$ & $-5.23$ & $-6.38$ \\
\hline
$E_B$ &$-2.272$ &$-2.356$ &$-2.357$ &$-2.370$  \\
\hline
\end{tabular}
\end{table}
The magnitudes of the $\Lambda p$ singlet and triplet scattering lengths are 
smaller than the
corresponding values of the Nijmegen NSC97e,f and J{\"u}lich '04 models
\cite{Hai05,Rij99}, which is also reflected in the small $\Lambda p$ cross section 
near threshold, see Figure \ref{fig:6.0}a. 
The mentioned models lead to a bound hypertriton \cite{Nog06,YNN2}. Although our $\Lambda p$
scattering lengths differ significantly from those of \cite{Hai05,Rij99}, the
$YN$ interaction based on chiral EFT also yields a correctly bound 
hypertriton, see Table \ref{tab:6.2}. 
Our singlet $\Sigma^+ p$ scattering length is about half as 
large as the values found for the $YN$ potentials in \cite{Hai05,Rij99}. Similar to those models and other 
$YN$ interactions, the value of the triplet $\Sigma^+ p$ scattering length is rather small. 
Contrary to \cite{Rij99}, but similar to \cite{Hai05} we found repulsion in this partial wave.

\begin{figure}[t]
\begin{center}
\resizebox{14.0cm}{16.47cm}{\includegraphics*[2.0cm,6.8cm][19.5cm,27cm]{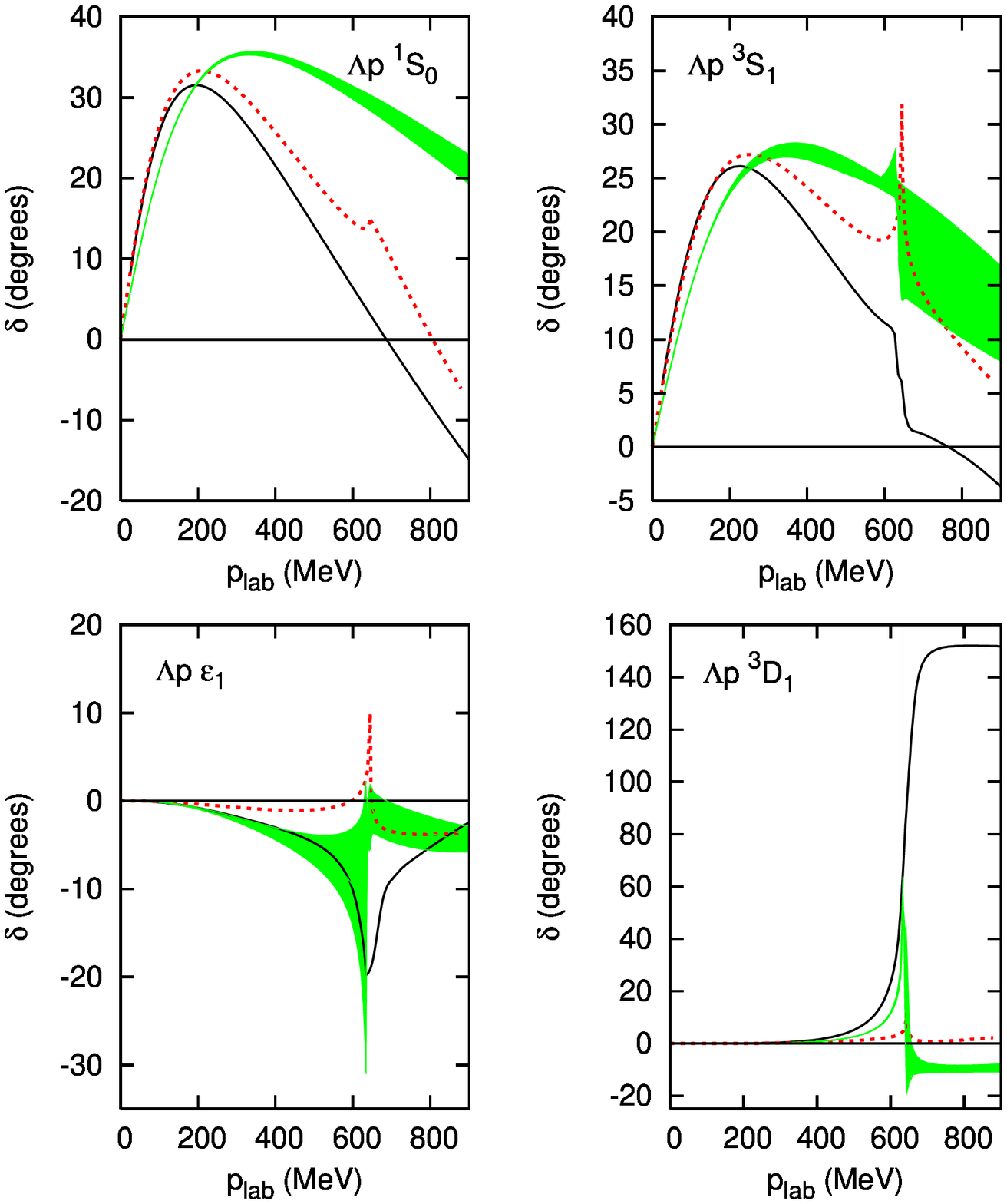}}
\end{center}
\caption{The $\Lambda p$ $S$-wave phase shifts $\delta$ as a function of $p_{{\rm lab}}$. The shaded band is the J{\"u}lich chiral EFT'06 A for $\Lambda = 550,...,700$ MeV, the dashed curve is the J{\"u}lich '04 model \cite{Hai05}, and the solid curve is the Nijmegen NSC97f model \cite{Rij99}. The Nijmegen $\epsilon_1$ phase shown here has an other sign convention than in Ref. \cite{Rij99}. Since the phases of the J{\"ulich '04 model are calculated in the isospin basis, their $\Sigma N$} threshold does not coincide with ours.}
\label{fig:6.4}
\end{figure}

\begin{figure}[h]
\begin{center}
\resizebox{14.0cm}{16.47cm}{\includegraphics*[2cm,6.8cm][19.5cm,27cm]{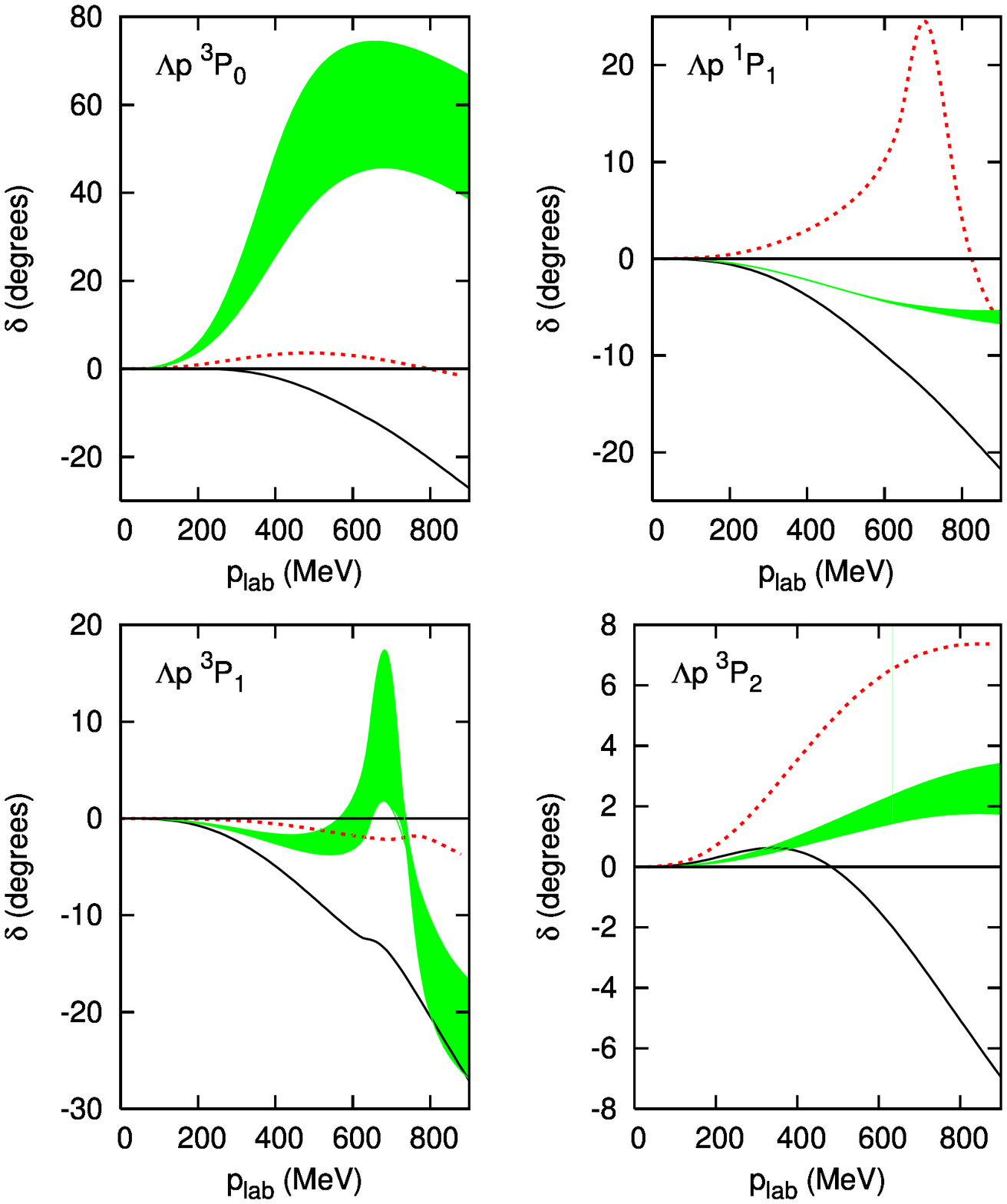}}
\end{center}
\caption{As in Figure \ref{fig:6.4}, but now for the $P$-wave phase shifts.}
\label{fig:6.5}
\end{figure}

\begin{figure}[h]
\begin{center}
\resizebox{14.0cm}{16.47cm}{\includegraphics*[2cm,6.8cm][19.5cm,27cm]{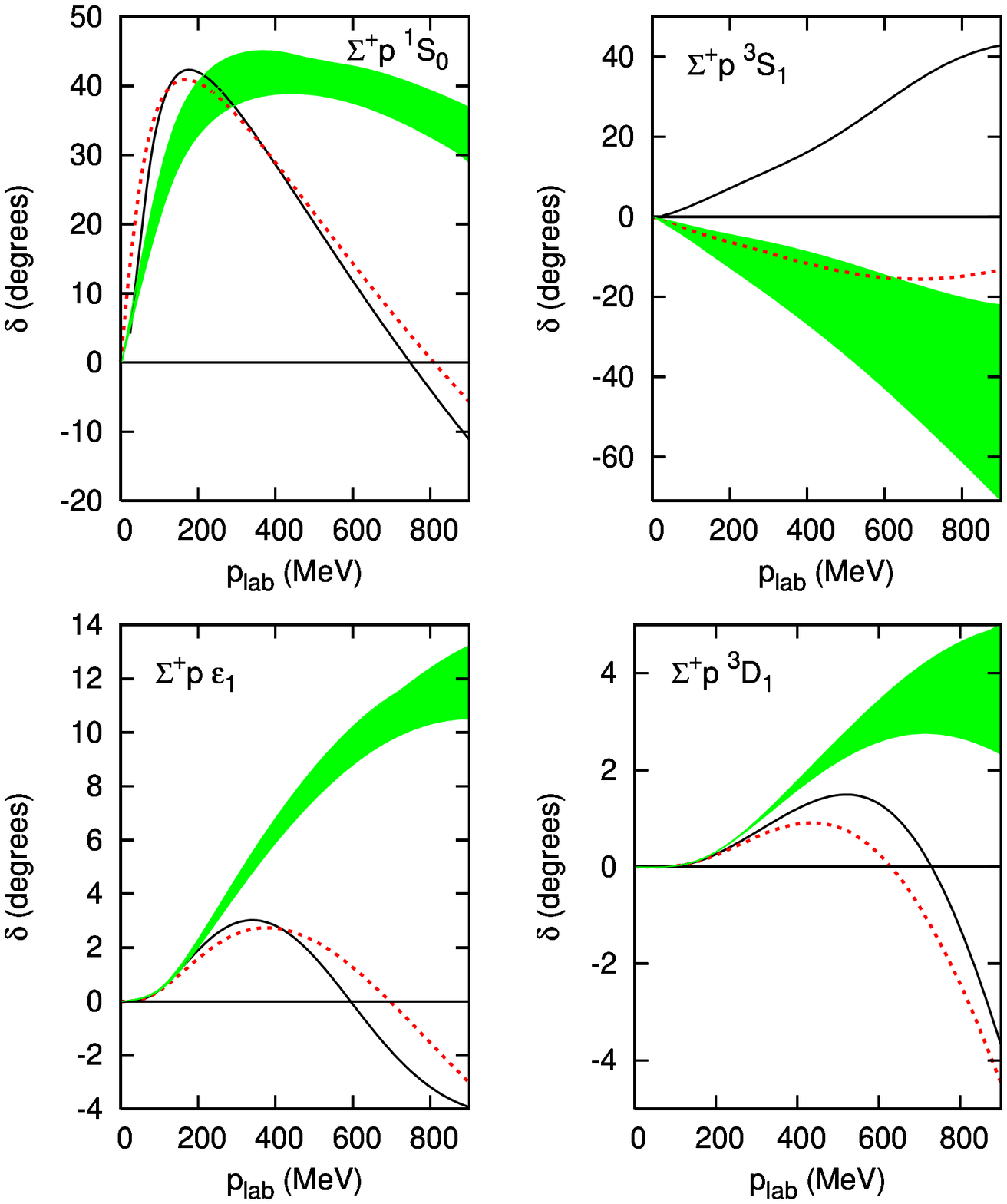}}
\end{center}
\caption{As in Figure \ref{fig:6.4}, but now for the $\Sigma^+ p$ $S$-wave phase shifts.}
\label{fig:6.6}
\end{figure}

\begin{figure}[h]
\begin{center}
\resizebox{14.0cm}{16.47cm}{\includegraphics*[2cm,6.8cm][19.5cm,27cm]{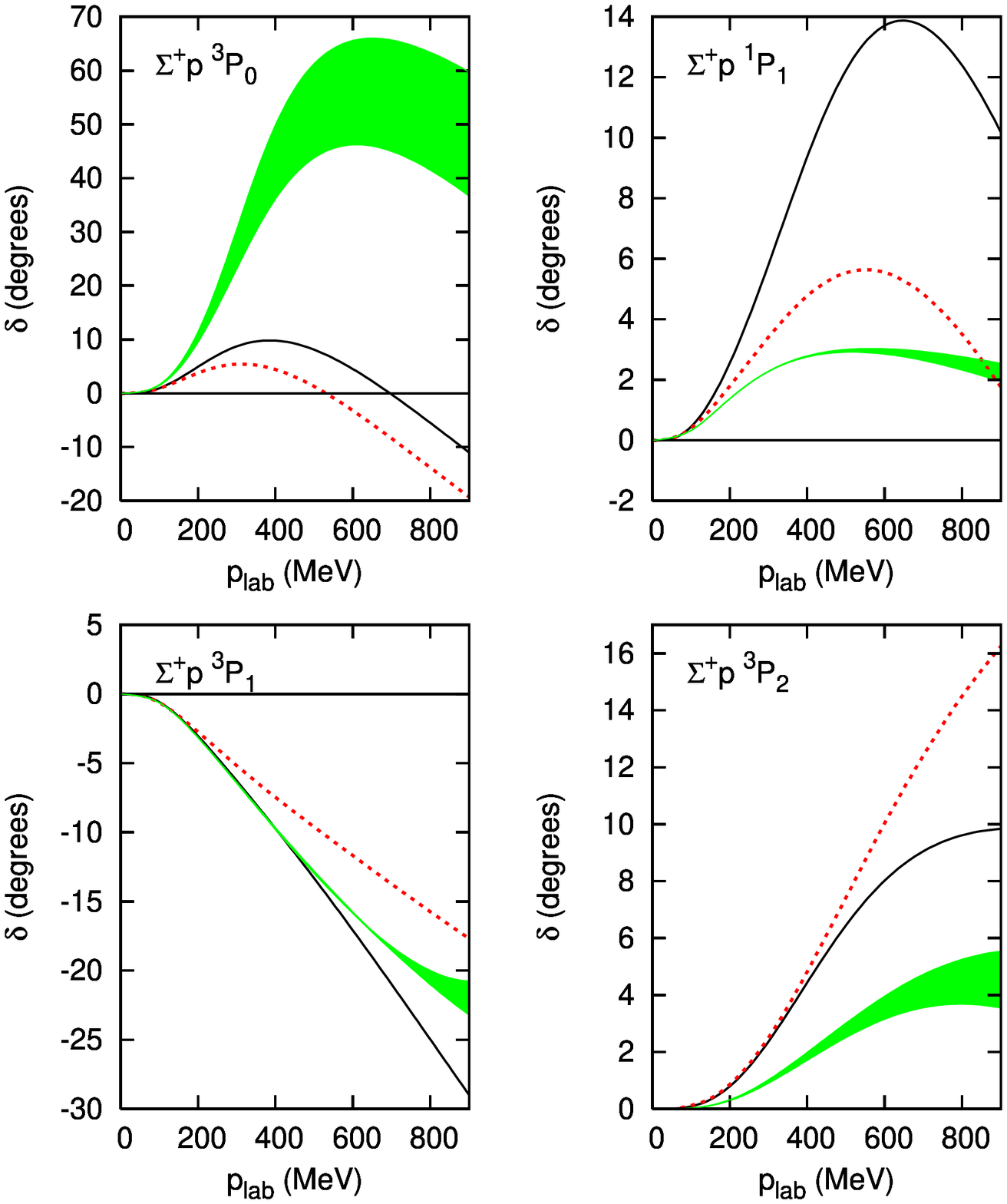}}
\end{center}
\caption{As in Figure \ref{fig:6.4}, but now for the $\Sigma^+ p$ $P$-wave phase shifts.}
\label{fig:6.7}
\end{figure}

The $S$- and $P$-wave phase shifts for $\Lambda p$ and $\Sigma^+p$ are shown in Figures \ref{fig:6.4} -- \ref{fig:6.7}. The shaded band represents the chiral EFT in the cut--off region $\Lambda=550,...,700$ MeV. As mentioned before, the 
limited accuracy of the $YN$ scattering data does not allow for a unique phase shift analysis. 
This explains why the chiral EFT phase shifts are quite different from the phase shifts of the models
presented in Refs.~\cite{Hai05,Rij99}. Actually, the predictions of the latter models also differ 
between each other in many partial waves. 
In both the $\Lambda p$ and $\Sigma^+ p$ ${}^1S_0$ and ${}^3P_0$ partial waves, the LO chiral EFT phase shifts are much larger at higher energies than the phases from \cite{Hai05,Rij99}. We emphasize that the empirical data, considered in the fitting procedure, are at lower energies. Also for the $NN$ interaction in leading order these partial waves were much larger than the Nijmegen phase shift analysis, see \cite{Epe00a}. It is expected that this problem for the $YN$ interaction can be solved by the derivative contact terms in a NLO calculation, just like in the $NN$ case. Our ${}^3S_1$ $\Sigma^+ p$ phase shift is repulsive like in \cite{Hai05}, but contrary to \cite{Rij99}. We remark that the $P$-waves are the result of pseudoscalar meson exchange only, since we only have contact terms in the $S$-waves. Contrary to \cite{Hai05}, there are no spin singlet to spin triplet transitions in the chiral EFT, because of the potential form in Eq. (\ref{eq:3.10}). Although the ${}^3D_1$ $\Lambda p$ phase shift near the $\Sigma N$ threshold rises quickly, it does not go through 90 degrees like in \cite{Rij99}. The opening of the $\Sigma N$ channel is also clearly seen in the ${}^3S_1$ $\Lambda p$ partial wave.

We have, so far, used the $SU(6)$ value for the pseudoscalar $F/(F+D)$ ratio;
$\alpha=0.4$. We studied the dependence on this parameter by
varying it within a range of 10 percent; after refitting the contact terms we
basically found an equally good description of the empirical data. Therefore,
we keep $\alpha$ to its $SU(6)$ value. As mentioned before, at NLO
one also has to consider symmetry breaking in the decay constants.


\section{Summary and outlook}
\label{chap:8}
In this paper we have studied the $YN$ interactions in a chiral effective
field theory approach based on a modified Weinberg power counting, analogous
to the $NN$ case in \cite{Epe05}. The symmetries of QCD are explicitly
incorporated. We assume that the $YN$ interactions are related via ${\rm
  SU(3)}_f$ symmetry. In principle the $YN$ interactions are also related to
the $NN$ interaction via ${\rm SU(3)}_f$ symmetry. However, since we have done our
study in leading order, in which the $NN$ interaction can not be described,
we do not consider the latter, but focus on the $YN$ interactions only.

The LO potential consists of two pieces: firstly, the longer-ranged 
one-pseudo{\-}scalar-meson exchanges, related via ${\rm SU(3)}_f$ symmetry in 
the well-known way and secondly, the shorter ranged four-baryon contact terms 
without derivatives. We have derived the ${\rm SU(3)}_f$ invariant four-baryon 
contact interaction. It contains five independent contact terms that need to
be determined from the empirical data. Contrary to the $NN$ case, the contact 
terms do not simply enter one specific partial wave because of the coupled 
particle channels and their ${\rm SU(3)}_f$ relations. Furthermore, a unique 
partial wave analysis for the $YN$ interaction does not exist, because of the 
scarce and inaccurate scattering data. Therefore we have directly fitted the 
parameters of the chiral EFT to the scattering observables.

The Lippmann-Schwinger equation for the LO chiral potential is solved in the partial wave basis. We have briefly discussed some details and have given the most general expressions for the $YN$ partial wave potentials in terms of the spinor invariants. The potential becomes unphysical for large momentum and has to be regularized. For this purpose we have multiplied the strong potential with an exponential regulator function. We used a cut--off in the range between $550$ and $700$ MeV. In order to incorporate the correct physical thresholds and the Coulomb interaction in the charged channels, we solve the Lippmann-Schwinger equation in the particle basis. The strong potential is, however, calculated in the isospin basis.

We have fitted the LO chiral EFT, with 5 free parameters, to 35 low-energy $YN$ scattering data. We obtained a good description of the empirical data, we found a total $\chi^2$ in the range between $27.8$ and $42.8$ for a cut--off in the range between $550$ and $700$ MeV. Also low-energy differential cross sections and higher energy cross sections, that were not included in the fitting procedure, were predicted quite well. Furthermore, the contact terms (found in the parameter search) are of natural size. As expected, in view of the inaccurate scattering data, the phase shifts
we found differ from those found obtained for conventional boson-exchange models. We remark that in LO only the $S$-waves contain contact terms, the other partial waves are parameter free. The ${}^1S_0$ and ${}^3P_0$ partial waves are too large at higher energies, this was also a problem in the LO $NN$ study \cite{Epe00a}. Probably this shortcoming will not occur in a NLO study, where derivative four-body contact terms may solve this problem.

We found that the chiral EFT yields a correctly bound hypertriton \cite{Nog06}. We did not explicitly include the hypertriton binding energy in the fitting procedure, but we have fixed the relative strength of the $\Lambda N$ singlet and triplet $S$-waves in such a way that a bound hypertriton could be obtained. We found that a $\Lambda p$ singlet scattering length of $-1.8$ fm leads to the correct binding energy. 

Our findings show that the chiral effective field theory scheme, applied in Ref.~\cite{Epe05} to the $NN$ interaction, 
also works well for the $YN$ interaction. In the future it will be interesting to study the convergence 
of the chiral EFT for the $YN$ interaction by doing NLO and NNLO calculations. In view of 
hypernucleus calculations, three baryon forces that naturally arise in chiral EFT, should be 
investigated too. Also a combined $NN$ and $YN$ study in chiral EFT, 
starting with a NLO calculation, needs to be performed. Work in this direction is in progress.

\ack
We are very grateful to Andreas Nogga for providing us with the values for the
hypertriton binding energy. We acknowledge useful discussions with Christoph Hanhart.
This research is part of the EU Integrated Infrastructure Initiative Hadron Physics Project
under contract number RII3-CT-2004-506078. Work supported in part by DFG (SFB/TR 16,
``Subnuclear Structure of Matter'').

\appendix
\section{Fierz theorem}
\label{app:A}
\countzero
Given the elements of the Clifford algebra, which are $4\times 4$-matrices,
\begin{equation}
\Gamma_1=1~, ~~\Gamma_2=\gamma^\mu~,~~ \Gamma_3=\sigma^{\mu\nu}~, ~~
\Gamma_4=\gamma^\mu\gamma_5~, ~~\Gamma_5=\gamma_5 ~ ,
\label{eq:A.1}
\end{equation}
the Fierz theorem \cite{Che84} tells us that
\begin{equation}
\sum_i\ C_i \left(\Gamma_i\right)_{ab}\left(\Gamma_i\right)_{cd}=\sum_k\ \tilde{C}_k \left(\Gamma_k\right)_{ad}\left(\Gamma_k\right)_{cb}\ ,
\end{equation}
where the new coefficients $\tilde{C}_k$ are related to the old coefficients $C_i$ by
\begin{eqnarray}
\left(
\begin{array}{l}
\tilde{C}_1 \\
\tilde{C}_2 \\
\tilde{C}_3 \\
\tilde{C}_4 \\
\tilde{C}_5 \\
\end{array}
\right) 
&=&\frac{1}{4}
\left(
\begin{array}{rrrrr}
1           & 4 &12 &-4 & 1 \\
1           &-2 & 0 &-2 &-1 \\
\frac{1}{2} & 0 &-2 & 0 &\frac{1}{2} \\
-1          &-2 & 0 &-2 & 1 \\
1           &-4 &12 & 4 & 1 
\end{array}
\right) 
\left(
\begin{array}{l}
C_1 \\
C_2 \\
C_3 \\
C_4 \\
C_5
\end{array}
\right) 
\ .
\label{eq:A.2}
\end{eqnarray}

\section{Partial wave projection}
\label{app:B}
\countzero
Because of rotational invariance and parity conservation, the potential 
can be expanded into the following set of 8 spinor invariants, see for example \cite{SNRV71,MRS89}. Introducing
\begin{equation}
  {\bf q}=\frac{1}{2}({\bf p}_{f}+{\bf p}_{i})\ , \
  {\bf k}={\bf p}_{f}-{\bf p}_{i}\ , \           
  {\bf n}={\bf p}_{i}\times {\bf p}_{f}\ ,
\label{eq:B.1} 
\end{equation}
we choose for the operators $P_{i}$ in spin-space
\begin{equation}
\begin{array}{ll}
  P_{1}=1\ ,  &  P_{2}= 
 \mbox{\boldmath $\sigma$}_1\cdot\mbox{\boldmath $\sigma$}_2 \ ,\\[0.2cm]
 P_{3}=(\mbox{\boldmath $\sigma$}_1\cdot{\bf k})(\mbox{\boldmath $\sigma$}_2\cdot{\bf k})
 -\frac{1}{3}(\mbox{\boldmath $\sigma$}_1\cdot\mbox{\boldmath $\sigma$}_2) 
  {\bf k}^2\ , & P_{4}=\frac{i}{2}(\mbox{\boldmath $\sigma$}_1+
 \mbox{\boldmath $\sigma$}_2)\cdot{\bf n} \ ,\\[0.2cm]
 P_{5}=(\mbox{\boldmath $\sigma$}_1\cdot{\bf n})(\mbox{\boldmath $\sigma$}_2\cdot{\bf n})\ ,
  & P_{6}=\frac{i}{2}(\mbox{\boldmath $\sigma$}_1-
 \mbox{\boldmath $\sigma$}_2)\cdot{\bf n} \ ,\\[0.2cm]
 \multicolumn{2}{l}{
 P_{7}=(\mbox{\boldmath $\sigma$}_1\cdot{\bf q})(\mbox{\boldmath $\sigma$}_2\cdot{\bf k})
 +(\mbox{\boldmath $\sigma$}_1\cdot{\bf k})(\mbox{\boldmath $\sigma$}_2\cdot{\bf q})\ ,
 } \\[0.2cm]
 \multicolumn{2}{l}{
 P_{8}=(\mbox{\boldmath $\sigma$}_1\cdot{\bf q})(\mbox{\boldmath $\sigma$}_2\cdot{\bf k})
 -(\mbox{\boldmath $\sigma$}_1\cdot{\bf k})(\mbox{\boldmath $\sigma$}_2\cdot{\bf q})\ .
 }
\end{array}
\label{eq:B.2} 
\end{equation}
Here we follow \cite{MRS89}, where in contrast to \cite{NRS78}, we have chosen $P_{3}$ to be a purely `tensor-force' operator. The operators $P_3$, $P_5$ and $P_7$ give rise to triplet coupled states (${}^3S_1\leftrightarrow {}^3D_1$, etc.). The operators $P_6$ and $P_8$ give spin singlet-triplet transitions (${}^1P_1\leftrightarrow {}^3P_1$, etc.). The expansion of the potential in spinor-invariants reads
\begin{equation}
 V({\bf p}_f,{\bf p}_i) = \sum_{i=1}^8\ V^{(i)}({\bf p}_f,{\bf p}_i)\ P_i({\bf p}_f,{\bf p}_i)\ .
\label{eq:B.3} 
\end{equation}

We will use the following shorthand notation for the potentials in the LSJ basis for the two parity ($P$) classes: \\
(i) $P=(-)^{J}$:
\begin{eqnarray}
     V^{J}_{0,0}  =  \left(J0J\left|V\right|J0J\right)
     & \hspace*{0.5cm} , \hspace*{0.5cm} &
     V^{J}_{0,2}  =  \left(J0J\left|V\right|J1J\right) \ ,\nonumber \\[0.2cm]
     V^{J}_{2,0}  =  \left(J1J\left|V\right|J0J\right)
     & \hspace*{0.5cm} , \hspace*{0.5cm} &
     V^{J}_{2,2}  =  \left(J1J\left|V\right|J1J\right)\ .
\end{eqnarray}
(ii) $P=-(-)^{J}$:
\begin{eqnarray}
     V^{J}_{1,1}  =  \left(J-1,1J\left|V\right|J-1,1J\right)
     & \hspace*{0.25cm} , \hspace*{0.25cm} &
     V^{J}_{1,3}  =   \left(J-1,1J\left|V\right|J+1,1J\right)\ , \nonumber \\[0.2cm]
     V^{J}_{3,1}  =  \left(J+1,1J\left|V\right|J-1,1J\right)
     & \hspace*{0.25cm} , \hspace*{0.25cm} &
     V^{J}_{3,3}  =  \left(J+1,1J\left|V\right|J+1,1J\right)\ ,
\end{eqnarray}
where it is always understood that the final and initial state
momenta are $p_{f}$ and $p_{i}$ respectively, e.g. $V^{J}_{0,0}= V^{J}_{0,0}(p_{f},p_{i})$ etc. 
Using the nomenclature $V_J^{(1)}=V_J^{(C)}$ for the central potential, $V_J^{(2)}=V_J^{(\sigma )}$ for the spin--spin potential, 
$V_J^{(3)}=V_J^{(T)}$ for the tensor potential, $V_J^{(4)}=V_J^{(SO)}$ for the spin--orbit potential, $V_J^{(5)}=V_J^{(Q)}$ for the quadratic spin--orbit potential and $V_J^{(6)}=V_J^{(ASO)}$ for the antisymmetric spin--orbit potential, the following partial wave potentials are found for $J>0$.
\begin{eqnarray}
V^J_{0,0}&=&4\pi\left[V_J^{(C)}-3V_J^{(\sigma)}+p_f^2p_i^2\left(e_{0,0}^{(5,+)}V_{J-2}^{(Q)}+f_{0,0}^{(5,+)}V_J^{(Q)}+g_{0,0}^{(5,+)}V_{J+2}^{(Q)}\right) 
\right. \nonumber \\ && \left. 
+\left(p_f^2+p_i^2\right)\cos 2\psi \ V_J^{(7)} \right] \ ,\nonumber \\
V^J_{0,2}&=&4\pi\frac{\sqrt{J(J+1)}}{2J+1}\left[p_fp_i\left(V_{J-1}^{(ASO)}-V_{J+1}^{(ASO)}\right)+
2p_fp_i\left(V_{J-1}^{(8)}-V_{J+1}^{(8)}\right)\right] \ ,\nonumber \\
V^J_{2,0}&=&4\pi\frac{\sqrt{J(J+1)}}{2J+1}\left[p_fp_i\left(V_{J-1}^{(ASO)}-V_{J+1}^{(ASO)}\right)-
2p_fp_i\left(V_{J-1}^{(8)}-V_{J+1}^{(8)}\right)\right]\ ,\nonumber \\
V^J_{2,2}&=&4\pi\left[V_J^{(C)}+V_J^{(\sigma)}+\frac{2}{3}\left(p_f^2+p_i^2\right)\left(V_J^{(T)}-\frac{1}{2}\sin 2\psi \ \left\{\frac{2J+3}{2J+1}V_{J-1}^{(T)}+\frac{2J-1}{2J+1}V_{J+1}^{(T)}\right\}\right)
\right. \nonumber \\ && \left. 
-p_fp_i\frac{1}{2J+1}\left(V_{J-1}^{(SO)}-V_{J+1}^{(SO)}\right)
+p_f^2p_i^2\left(e_{1,1}^{(5,+)}V_{J-2}^{(Q)}+f_{1,1}^{(5,+)}V_J^{(Q)}+g_{1,1}^{(5,+)}V_{J+2}^{(Q)}\right) 
\right. \nonumber \\ && \left. 
-\left(p_f^2+p_i^2\right)\cos 2\psi \ V_J^{(7)} \right] \ ,\nonumber \\
V^J_{1,1}&=&4\pi\left[V_{J-1}^{(C)}+V_{J-1}^{(\sigma)}+\frac{2}{3}\left(p_f^2+p_i^2\right)\frac{J-1}{2J+1}\left(-V_{J-1}^{(T)}+\frac{1}{2}\sin 2\psi \ \left\{\frac{2J-3}{2J-1}V_{J}^{(T)}+
\right.\right.\right. \nonumber \\ && \left.\left.\left.
\frac{2J+1}{2J-1}V_{J-2}^{(T)}\right\}\right)
+p_fp_i\frac{J-1}{2J-1}\left(V_{J-2}^{(SO)}-V_{J}^{(SO)}\right)+
p_f^2p_i^2\left(e_{J-1,J-1}^{(5,-)}V_{J-3}^{(Q)}+
\right.\right. \nonumber \\ &&\left.\left.
f_{J-1,J-1}^{(5,-)}V_{J-1}^{(Q)}+g_{J-1,J-1}^{(5,-)}V_{J+1}^{(Q)}\right) 
-\left(p_f^2+p_i^2\right)\cos 2\psi \ \frac{1}{2J+1}V_{J-1}^{(7)} 
\right] \ ,\nonumber \\
V^J_{1,3}&=&4\pi\left[2\left(p_f^2+p_i^2\right)\frac{\sqrt{J(J+1)}}{2J+1}\left(\sin 2\psi \ V_J^{(T)}- \left\{\cos^2\psi \ V_{J-1}^{(T)}+\sin^2\psi \ V_{J+1}^{(T)}\right\}\right)
\right. \nonumber \\ && \left.
-p_f^2p_i^2f_{J+1,J-1}^{(5,-)}\left(V_{J+1}^{(Q)}-V_{J-1}^{Q)}\right) + \left(p_f^2+p_i^2\right)\frac{2\sqrt{J(J+1)}}{2J+1}\left(-\sin^2 \psi \ V_{J+1}^{(7)}+
\right.\right. \nonumber \\ && \left.\left.
\cos^2 \psi \ V_{J-1}^{(7)}\right)
\vphantom{\frac{\sqrt{J(J+1)}}{2J+1}}\right] \ ,\nonumber \\
V^J_{3,1}&=&4\pi\left[2\left(p_f^2+p_i^2\right)\frac{\sqrt{J(J+1)}}{2J+1}\left(\sin 2\psi \ V_J^{(T)}- \left\{\cos^2\psi \ V_{J+1}^{(T)}+\sin^2\psi \ V_{J-1}^{(T)}\right\}\right)
\right. \nonumber \\ && \left.
-p_f^2p_i^2f_{J+1,J-1}^{(5,-)}\left(V_{J+1}^{(Q)}-V_{J-1}^{Q)}\right) + \left(p_f^2+p_i^2\right)\frac{2\sqrt{J(J+1)}}{2J+1}\left(-\sin^2 \psi \ V_{J-1}^{(7)}+
\right.\right. \nonumber \\ && \left.\left.
\cos^2 \psi \ V_{J+1}^{(7)}\right)
\vphantom{\frac{\sqrt{J(J+1)}}{2J+1}}\right]\ ,\nonumber \\
V^J_{3,3}&=&4\pi\left[V_{J+1}^{(C)}+V_{J+1}^{(\sigma)}+\frac{2}{3}\left(p_f^2+p_i^2\right)\frac{J+2}{2J+1}\left(-V_{J+1}^{(T)}+\frac{1}{2}\sin 2\psi \ \left\{\frac{2J+5}{2J+3}V_{J}^{(T)}+
\right.\right.\right. \nonumber \\ && \left.\left.\left.
\frac{2J+1}{2J+3}V_{J+2}^{(T)}\right\}\right)
-p_fp_i\frac{J+2}{2J+3}\left(V_{J}^{(SO)}-V_{J+2}^{(SO)}\right)+
p_f^2p_i^2\left(e_{J+1,J+1}^{(5,-)}V_{J-1}^{(Q)}+
\right.\right. \nonumber \\ &&\left.\left.
f_{J+1,J+1}^{(5,-)}V_{J+1}^{(Q)}+g_{J+1,J+1}^{(5,-)}V_{J+3}^{(Q)}\right) 
+\left(p_f^2+p_i^2\right)\cos 2\psi \ \frac{1}{2J+1}V_{J+1}^{(7)} 
\right] \ .
\end{eqnarray}
For $J=0$ the two non-zero partial wave potentials are
\begin{eqnarray}
V^J_{0,0}&=&4\pi\left[V_0^{(C)}-3V_0^{(\sigma)}+\frac{2}{3}p_f^2p_i^2\left(-V_0^{(Q)}+V_{2}^{(Q)}\right) 
+\left(p_f^2+p_i^2\right)\cos 2\psi \ V_0^{(7)} \right] \ ,\nonumber \\
V^J_{3,3}&=&4\pi\left[V_{1}^{(C)}+V_{1}^{(\sigma)}+\frac{4}{3}\left(p_f^2+p_i^2\right)\left(-V_{1}^{(T)}+\frac{1}{2}\sin 2\psi \ \left\{\frac{5}{3}V_{0}^{(T)}+
\frac{1}{3}V_{2}^{(T)}\right\}\right)
\right. \nonumber \\ && \left. 
-p_fp_i\frac{2}{3}\left(V_{0}^{(SO)}-V_{2}^{(SO)}\right)+
\frac{2}{5}p_f^2p_i^2\left(
V_{1}^{(Q)}-V_{3}^{(Q)}\right) 
\right. \nonumber \\ && \left. 
+\left(p_f^2+p_i^2\right)\cos 2\psi \ V_{1}^{(7)} 
\vphantom{\frac{5}{3}}\right] \ .
\end{eqnarray}
In the formulae above we have used
\begin{eqnarray}
V^{(i)}_J(p_f,p_i)&=&\frac{1}{2}\int_{-1}^1d\cos\theta\ V^{(i)}({\bf p}_f,{\bf p}_i)P_J(\cos\theta)\ .
\end{eqnarray}
Details of the derivation and definitions of $\cos 2\psi$, $\sin 2\psi$ and the various $e^{(5)}$, $f^{(5)}$ and $g^{(5)}$ factors can be found in Appendix \ref{app:B.1}. We note that an additional overall (-) sign for the off-diagonal $V^J_{1,3}$ and $V^J_{3,1}$ has been used in the calculations.

\section{Partial wave projection of spinor invariants}
\label{app:B.1}
With the matrix elements for the spinor invariants in this appendix (found using the results of Appendix \ref{app:B.2}), the partial wave potentials in Appendix \ref{app:B} can be readily derived. The derivation in this appendix is an extension of the derivation for the $NN$ case in \cite{Rij02}.

Distinguishing between the partial waves with parity $P=(-)^{J}$
and $P=-(-)^{J}$, we write the potential matrix elements on the
LSJ-basis in the following way (see {\it e.g.}\ \cite{SNRV71}): \\
(i) $P=(-)^{J}$:
\begin{equation}
  (p_f ; L' S' J' M'|\ V\ |p_i; L S J M) =  4\pi\, \delta_{J^{\prime}J}\ \delta_{M^{\prime}M}\ \delta_{L^{\prime}L}\, V^{J,+}(S',S)\ .
\label{eq:9.1}  \end{equation}
(ii) $P=-(-)^{J}$:
\begin{equation}
  (p_f ; L' S' J' M'|\ V\ |p_i; L S J M) =  4\pi\,
\delta_{J^{\prime}J}\,\delta_{M^{\prime}M}\,\delta_{S^{\prime}S}\,
 V^{J,-}(L',L)\ . 
\label{eq:9.2}  \end{equation}
For notational convenience we will use as an index the parity factor
$\eta$, which is defined by writing $ P=\eta (-)^{J}$.
The $P=(-)^{J}$ states contain the spin singlet and triplet-uncoupled
states ($\eta=+$), and the $P=-(-)^{J}$ states contain
the spin triplet-coupled states ($\eta=-$).
 
\noindent Below we list the partial wave matrix elements for $ \eta = \pm$ for
the different $V^{(i)}\ P_{i}, (i =1,...,8)$. Here we
restrict ourselves to the matrix elements $\neq 0$.  \\
1. {\it central} $P_{1}=1$:
\begin{eqnarray}
     (p_f ; L' S' J' M' | V^{(1)}P_{1} |p_i; L S J M) &=&
4\pi\,  \delta_{J^{\prime}J}\,\delta_{M^{\prime}M}\,
F^{J,\eta}_{1}(L^{\prime}\ S',L\ S)\ ,
    \label{PWP1} \\[0.3cm]
{\rm with } \hspace*{0.5cm}  F^{J,\eta}_{1}(L^{\prime}\ S',L\ S)\ &=&
  \delta_{L^{\prime}L}\, \delta_{S^{\prime}S}\, V^{(1)}_{L}\ . \nonumber
\end{eqnarray}
2. {\it spin-spin} $P_{2}=
\mbox{\boldmath $\sigma$}_1\cdot\mbox{\boldmath $\sigma$}_2$:
\begin{eqnarray}
     (p_f ; L' S' J' M' | V^{(2)}P_{2} |p_i; L S J M) &=&
4\pi\,  \delta_{J^{\prime}J}\,\delta_{M^{\prime}M}\,
F^{J,\eta}_{2}(L^{\prime}\ S',L\ S)\ ,
\label{PWP2} \\[0.3cm]
{\rm with} \hspace*{0.5cm}  F^{J,\eta}_{2}(L^{\prime}\ S',L\ S)\ &=&
   \delta_{L^{\prime}L}\, \delta_{S^{\prime}S}\,
   \left[2S(S+1)-3\right] V^{(2)}_{L}\ .
\nonumber \end{eqnarray}
3. {\it tensor} $ P_{3}= (\mbox{\boldmath $\sigma$}_1\cdot{\bf k})
(\mbox{\boldmath $\sigma$}_2\cdot{\bf k}) - \frac{1}{3} 
(\mbox{\boldmath $\sigma$}_1\cdot\mbox{\boldmath $\sigma$}_2) {\bf k}^{2}$:
\begin{equation}
     (p_f ; L' S' J' M' | V^{(3)}P_{3} |p_i; L S J M) =
\frac{8\pi}{3} (p_f^{2}+p_i^{2})\,
\delta_{J^{\prime}J}\,\delta_{M^{\prime}M}\, F^{J,\eta}_{3}(i,j)\ ,
\label{eq:9.3}  \end{equation}
where $i=S'$ and $j=S$ for $\eta=+$, respectively $i=L'$ and $j=L$
for $\eta=-$. \\
(i) triplet uncoupled: $L=L'=J,\ S=S'=1$
\begin{eqnarray}
 F^{J,+}_{3}(1,1)&=&
 \left[ V^{(3)}_{J}- \frac{1}{2}\sin 2\psi
 \left(\frac{2J+3}{2J+1}V^{(3)}_{J-1}+
 \frac{2J-1}{2J+1}V^{(3)}_{J+1}\right)\right]\ . \end{eqnarray}
(ii) triplet coupled: $L=J\pm 1,\ L'=J\pm 1,\ S=S'=1$
 \begin{eqnarray}
  F^{J,-}_{3}(J-1,J-1) &=& \frac{J-1}{2J+1}\ \left[ -V^{(3)}_{J-1}
  + \frac{1}{2} \sin 2\psi 
  \left( \frac{2J-3}{2J-1} V^{(3)}_{J}
  + \frac{2J+1}{2J-1} V^{(3)}_{J-2} \right) \right]\ ,
\nonumber
\\
  F^{J,-}_{3}(J-1,J+1) &=& -3 \frac{\sqrt{J(J+1)}}{2J+1}\ \left[
  - \sin 2\psi\ V^{(3)}_{J} 
 +\left(\cos^{2}\psi V^{(3)}_{J-1}
 +\sin^{2}\psi V^{(3)}_{J+1}\right) \right]\ ,
\nonumber
\\
  F^{J,-}_{3}(J+1,J-1) &=& -3 \frac{\sqrt{J(J+1)}}{2J+1}\ \left[
  - \sin 2\psi\ V^{(3)}_{J}  
 +\left(\sin^{2}\psi V^{(3)}_{J-1}
 +\cos^{2}\psi V^{(3)}_{J+1}\right) \right]\ ,
\nonumber
\\
  F^{J,-}_{3}(J+1,J+1) &=& \frac{J+2}{2J+1}\ \left[ -V^{(3)}_{J+1}
  + \frac{1}{2} \sin 2\psi 
  \left( \frac{2J+5}{2J+3} V^{(3)}_{J}
  + \frac{2J+1}{2J+3} V^{(3)}_{J+2} \right) \right]\ ,
\end{eqnarray}
where we introduced
\begin{equation}
\cos\psi = \frac{p_i}{\sqrt{p_f^{2}+ p_i^{2}}} \hspace{0.5cm} , \hspace{0.5cm}
\sin\psi = \frac{p_f}{\sqrt{p_f^{2}+ p_i^{2}}}\ .
\label{eq:9.4}  \end{equation}
4. {\it spin-orbit} $ P_{4}=\frac{i}{2}
(\mbox{\boldmath $\sigma$}_1+\mbox{\boldmath $\sigma$}_2)\cdot{\bf n}$:
\begin{eqnarray}
     (p_f ; L' S' J' M'| V^{(4)}P_{4} |p_i; L S J M)&=&
      4\pi\, p_f p_i \delta_{J^{\prime}J}\,\delta_{M^{\prime}M}\,
  F^{J,\eta}_{4}(i,j)\ .
\label{eq:9.5}  \end{eqnarray}
(i) triplet uncoupled: $L=L'=J,\ S=S'=1$
\begin{equation}
  F^{J,+}_{4}(1,1)= - \left(
         V^{(4)}_{J-1}-V^{(4)}_{J+1} \right)/(2J+1)\ .
\label{eq:9.6}  \end{equation}
(ii) triplet coupled: $L=J\pm 1,\ L'=J\pm 1,\ S=S'=1$
\begin{eqnarray}
   F^{J,-}_{4}(J-1,J-1)&=& \hspace*{0.3cm}
    \frac{(J-1)}{(2J-1)}
    \left(V^{(4)}_{J-2}-V^{(4)}_{J}\right) \ ,\nonumber \\[0.2cm]
   F^{J,-}_{4}(J+1,J+1)&=& -
    \frac{(J+2)}{(2J+3)}\left(V^{(4)}_{J}-V^{(4)}_{J+2}\right)\ .
\label{eq:9.7}  \end{eqnarray}
5. {\it quadratic-spin-orbit} $ P_{5}= (\mbox{\boldmath $\sigma$}_1\cdot{\bf n})(\mbox{\boldmath $\sigma$}_2\cdot{\bf n})$:
\begin{equation}
     (p_f ; L' S' J' M' | V^{(5)}P_{5} |p_i; L S J M) =
4\pi p_f^{2}p_i^{2}\,
\delta_{J^{\prime}J}\,\delta_{M^{\prime}M}\, F^{J,\eta}_{5}(i,j)\ .
\label{eq:99.3}  \end{equation}
(i) singlet: $L=L'=J,\ S=S'=0$
\begin{eqnarray}
 F^{J,+}_{5}(0,0)&=&
  e^{(5,+)}_{0,0}V^{(5)}_{J-2}+f^{(5,+)}_{0,0}V^{(5)}_{J}+g^{(5,+)}_{0,0}V^{(5)}_{J+2}\ .
\end{eqnarray}
(ii) triplet uncoupled: $L=L'=J,\ S=S'=1$
\begin{eqnarray}
 F^{J,+}_{5}(1,1)&=&
  e^{(5,+)}_{1,1}V^{(5)}_{J-2}+ f^{(5,+)}_{1,1}V^{(5)}_{J}+g^{(5,+)}_{1,1}V^{(5)}_{J+2}\ ,
\end{eqnarray}
where we introduced
\begin{eqnarray}
e^{(5,+)}_{0,0}&=&+\frac{J(J-1)}{(2J-1)(2J+1)} \hspace{0.5cm} , \hspace{0.5cm}
e^{(5,+)}_{1,1}=+\frac{(J-1)(J+2)}{(2J-1)(2J+1)} \ ,\nonumber \\
f^{(5,+)}_{0,0}&=&-\frac{2(J^2+J-1)}{(2J-1)(2J+3)} \hspace{0.5cm} , \hspace{0.5cm}
f^{(5,+)}_{1,1}=-\frac{2(J-1)(J+2)}{(2J-1)(2J+3)} \ ,\nonumber \\
g^{(5,+)}_{0,0}&=&+\frac{(J+1)(J+2)}{(2J+1)(2J+3)} \hspace{0.5cm} , \hspace{0.5cm}
g^{(5,+)}_{1,1}=+\frac{(J-1)(J+2)}{(2J+1)(2J+3)}\ .
 \end{eqnarray}
(iii) triplet coupled: $L=J\pm 1,\ L'=J\pm 1,\ S=S'=1$
 \begin{eqnarray}
  F^{J,-}_{5}(J-1,J-1) &=& e^{(5,-)}_{J-1,J-1}V^{(5)}_{J-3}+f^{(5,-)}_{J-1,J-1}V^{(5)}_{J-1}+g^{(5,-)}_{J-1,J-1}V^{(5)}_{J+1}\ ,
\nonumber\\
  F^{J,-}_{5}(J\pm 1,J\mp 1) &=&  -f^{(5,-)}_{J+1,J-1}\left[V^{(5)}_{J+1}-V^{(5)}_{J-1}\right]\ ,
\nonumber\\
  F^{J,-}_{5}(J+1,J+1) &=& e^{(5,-)}_{J+1,J+1}V^{(5)}_{J-1}+f^{(5,-)}_{J+1,J+1}V^{(5)}_{J+1}+g^{(5,-)}_{J+1,J+1}V^{(5)}_{J+3}\ ,
\end{eqnarray}
where we introduced
\begin{eqnarray}
e^{(5,-)}_{J-1,J-1}&=&-\frac{(J-1)(J-2)}{(2J-1)(2J-3)} \hspace{0.5cm} , \hspace{0.5cm}
e^{(5,-)}_{J+1,J+1}=-\frac{J(2J^2+7J+7)}{(2J+1)^2(2J+3)} \ ,\nonumber \\
f^{(5,-)}_{J-1,J-1}&=&2\frac{(2J^3-3J^2-2J+2)}{(2J+1)^2(2J-3)} \hspace{0.5cm} , \hspace{0.5cm}
f^{(5,-)}_{J+1,J+1}=2\frac{(2J^3+9J^2+10J+1)}{(2J+1)^2(2J+5)} \ ,\nonumber \\
g^{(5,-)}_{J-1,J-1}&=&-\frac{(2J^2-3J+2)(J+1)}{(2J+1)^2(2J-1)} \hspace{0.5cm} , \hspace{0.5cm}
g^{(5,-)}_{J+1,J+1}=-\frac{(J+2)(J+3)}{(2J+3)(2J+5)} \ ,\nonumber \\
f^{(5,-)}_{J+1,J-1}&=&2\frac{\sqrt{J(J+1)}}{(2J+1)^2}\ .
 \end{eqnarray}
6. {\it antisymmetric spin-orbit} $ P_{6}=\frac{i}{2}
(\mbox{\boldmath $\sigma$}_1-\mbox{\boldmath $\sigma$}_2)\cdot{\bf n}$:
\begin{eqnarray}
     (p_f ; L' S' J' M'| V^{(6)}P_{6} |p_i; L S J M)&=&
      4\pi\, p_f p_i \delta_{J^{\prime}J}\,\delta_{M^{\prime}M}\,
  F^{J,\eta}_{6}(i,j)\ .
\label{eq:99.5}  
\end{eqnarray}
(i) singlet-triplet uncoupled: $L=L'=J,\ S\neq S'$
\begin{equation}
  F^{J,+}_{6}(1,0)=F^{J,+}_{6}(0,1)= \frac{\sqrt{J(J+1)}}{2J+1} \left(
         V^{(6)}_{J-1}-V^{(6)}_{J+1} \right)\ .
\label{eq:99.6}  
\end{equation}
7. $ P_{7}= (\mbox{\boldmath $\sigma$}_1\cdot{\bf q})
(\mbox{\boldmath $\sigma$}_2\cdot{\bf k}) + (\mbox{\boldmath $\sigma$}_1\cdot{\bf k})(\mbox{\boldmath $\sigma$}_2\cdot{\bf q})$:
\begin{equation}
     (p_f ; L' S' J' M' | V^{(7)}P_{7} |p_i; L S J M) =
4\pi (p_f^{2}+p_i^{2})\,
\delta_{J^{\prime}J}\,\delta_{M^{\prime}M}\, F^{J,\eta}_{7}(i,j)\ .
\label{eq:9.3a}  \end{equation}
(i) singlet: $L=L'=J,\ S=S'=0$
\begin{eqnarray}
 F^{J,+}_{7}(0,0)&=&
 \cos 2\psi \ V^{(7)}_{J} \ .\end{eqnarray}
(ii) triplet uncoupled: $L=L'=J,\ S=S'=1$
\begin{eqnarray}
 F^{J,+}_{7}(1,1)&=&
 -\cos 2\psi \ V^{(7)}_{J} \ .\end{eqnarray}
(iii) triplet coupled: $L=J\pm 1,\ L'=J\pm 1,\ S=S'=1$
 \begin{eqnarray}
  F^{J,-}_{7}(J-1,J-1) &=& -\frac{1}{2J+1}\cos 2\psi \ V^{(7)}_{J-1}\ ,
\nonumber \\
  F^{J,-}_{7}(J-1,J+1) &=& 2 \frac{\sqrt{J(J+1)}}{2J+1}\ \left[-\sin^{2}\psi \ V^{(7)}_{J-1}+\cos^{2}\psi \ V^{(7)}_{J+1} \right]\ ,
\nonumber\\
  F^{J,-}_{7}(J+1,J-1) &=& 2 \frac{\sqrt{J(J+1)}}{2J+1}\ \left[-\sin^{2}\psi \ V^{(7)}_{J+1}+\cos^{2}\psi \ V^{(7)}_{J-1}\right]\ ,
\nonumber\\
  F^{J,-}_{7}(J+1,J+1) &=& \frac{1}{2J+1}\cos 2 \psi \ V^{(7)}_{J+1}\ .
\end{eqnarray}
8. $ P_{8}=(\mbox{\boldmath $\sigma$}_1\cdot{\bf q})
(\mbox{\boldmath $\sigma$}_2\cdot{\bf k}) - (\mbox{\boldmath $\sigma$}_1\cdot{\bf k})(\mbox{\boldmath $\sigma$}_2\cdot{\bf q})$:
\begin{eqnarray}
     (p_f ; L' S' J' M'| V^{(8)}P_{8} |p_i; L S J M)&=&
      4\pi\, \left(p_f^2 + p_i^2\right) \delta_{J^{\prime}J}\,\delta_{M^{\prime}M}\, F^{J,\eta}_{8}(i,j)\ .
\label{eq:99.5a}  
\end{eqnarray}
(i) singlet-triplet uncoupled: $L=L'=J,\ S\neq S'$
\begin{equation}
  F^{J,+}_{8}(1,0)=-F^{J,+}_{8}(0,1)= -\frac{\sqrt{J(J+1)}}{2J+1}\sin 2\psi \ \left(V^{(8)}_{J-1}-V^{(8)}_{J+1} \right)\ .
\label{eq:99.6a}  
\end{equation}

Henceforth, we will use the following shorthand notation for the potentials: \\
(i) $P=(-)^{J}$:
\begin{eqnarray}
     V^{J}_{0,0}  =  V^{J,+} (0,0)
     & \hspace*{0.5cm} , \hspace*{0.5cm} &
     V^{J}_{0,2}  =  V^{J,+} (0,1) \ ,\nonumber \\[0.2cm]
     V^{J}_{2,0}  =  V^{J,+} (1,0)
     & \hspace*{0.5cm} , \hspace*{0.5cm} &
     V^{J}_{2,2}  =  V^{J,+} (1,1)\ .
\label{eq:9.8}  \end{eqnarray}
(ii) $P=-(-)^{J}$:
\begin{eqnarray}
     V^{J}_{1,1}  =  V^{J,-} (J-1,J-1)
     & \hspace*{0.5cm} , \hspace*{0.5cm} &
     V^{J}_{1,3}  =  V^{J,-} (J-1,J+1) \ , \nonumber \\[0.2cm]
     V^{J}_{3,1}  =  V^{J,-} (J+1,J-1)
     & \hspace*{0.5cm} , \hspace*{0.5cm} &
     V^{J}_{3,3}  =  V^{J,-} (J+1,J+1)\ ,
\label{eq:9.9}  \end{eqnarray}
where it is always understood that the final and initial state
momenta are $p_{f}$ and $p_{i}$ respectively,
e.g. $V^{J}_{0,0}= V^{J}_{0,0}(p_{f},p_{i})$ etc.

\subsection{The LSJ representation operators}
\countzero
\label{app:B.2}
{}From the formulas given in this section the partial wave projections of the spinor invariants, as given above
, can be derived in a straightforward manner.

The spherical wave functions in momentum space with quantum numbers
J, L, S, are in the SYM-convention \cite{SYM57}
\begin{equation}
  {\mathcal Y}_{J L S}^{M}(\hat{\bf p})= i^{L}\
  \mbox{\large\bf $C$}^{J\ L\ S}_{M\ m\ \mu}
  Y^{L}_{m}(\hat{\bf p}) \chi^{S}_{\mu} \ ,
\label{eq:appb.1} \end{equation}
where $\chi$ is the two-nucleon spin wave function. Then
\begin{eqnarray}
    \left({\bf S}\cdot\hat{\bf p}\right) {\mathcal Y}_{J L S}^{M}(\hat{\bf p})&=&
 i\sqrt{6}\ \left\{ \sqrt{\frac{L}{2L-1}}
 \left[\begin{array}{ccc}
       L & S & J \\
       1 & 1 & 0 \\
     L-1 & S & J
        \end{array}\right] {\mathcal Y}_{J L-1 S}^{M}(\hat{\bf p}) \right.
  \nonumber \\ & & \nonumber \\ & & \left. 
  + \sqrt{\frac{L+1}{2L+3}}
 \left[\begin{array}{ccc}
       L & S & J \\
       1 & 1 & 0 \\
     L+1 & S & J
        \end{array}\right] {\mathcal Y}_{J L+1 S}^{M}(\hat{\bf p}) \right\}\ ,
\nonumber \\ 
\label{eq:appb.2} \end{eqnarray}
where ${\bf S}=\left(\mbox{\boldmath $\sigma$}_1+\mbox{\boldmath $\sigma$}_2\right)/2$. The $9j$-symbols differ from \cite{Edm57}, formula
$(6.4.4)$, in the replacement of the $3j$-symbols by the
Clebsch-Gordan coefficients and by leaving out the $m_{33}$--summation.
 Working this out explicitly, we find
\begin{eqnarray}
    \left({\bf S}\cdot\hat{\bf p}\right) {\mathcal Y}_{J J-1 1}^{M}(\hat{\bf p})&=&
    -i\ a_{J}\ {\mathcal Y}_{J J 1}^{M}(\hat{\bf p}) \ ,\nonumber \\
    \left({\bf S}\cdot\hat{\bf p}\right) {\mathcal Y}_{J J+1 1}^{M}(\hat{\bf p})&=&
  \hspace*{0.3cm} i\ b_{J}\ {\mathcal Y}_{J J 1}^{M}(\hat{\bf p}) \ ,\nonumber \\
    \left({\bf S}\cdot\hat{\bf p}\right) {\mathcal Y}_{J\ J\ 1}^{M}(\hat{\bf p})&=&
  \hspace*{0.3cm}  i\ a_{J}\ {\mathcal Y}_{J J-1 1}^{M}(\hat{\bf p}) -i\
    b_{J}\ {\mathcal Y}_{J J+1 1}^{M}(\hat{\bf p})\ ,          
\label{eq:appb.3} \end{eqnarray}
where
\begin{equation}
  a_{J}= -\sqrt{\frac{J+1}{2J+1}} \hspace*{1.0cm} ,
  \hspace*{1.0cm} b_{J}= -\sqrt{\frac{J}{2J+1}} \ .
\label{eq:appb.4} \end{equation}
 Ordering the states according to $L= J-1, L=J, L=J+1$ ,
we can write in matrix form \\
\begin{eqnarray}
  \left(\begin{array}{ccc}
   L&=&J-1 \\  & &J \\  & & J+1 \end{array}\right\|
  {\bf S}\cdot\hat{\bf p}
  \left\|\begin{array}{ccc}
  L&=& J-1 \\ & & J \\ & & J+1 \end{array}\right) =
  \left(\begin{array}{ccc}
  0 & i a_{J} & 0 \\ -i a_{J} & 0 & i b_{J} \\ 0 & -i b_{J} & 0
  \end{array} \right) \ .
\label{eq:appb.5} \end{eqnarray}
Similarly we find for the operator ${\bf AS}=\left(\mbox{\boldmath $\sigma$}_1-\mbox{\boldmath $\sigma$}_2\right)/2$:
\begin{eqnarray}
    \left({\bf AS}\cdot\hat{\bf p}\right) {\mathcal Y}_{J L S}^{M}(\hat{\bf p})&=&
 \sum_{S'}i\sqrt{3}\ \left\{ \sqrt{\frac{L}{2L-1}}
 \left[\begin{array}{ccc}
       L & S & J \\
       1 & 1 & 0 \\
     L-1 & S' & J
        \end{array}\right] {\mathcal Y}_{J L-1 S'}^{M}(\hat{\bf p}) +\right.
  \nonumber \\ & & \nonumber \\ & & \left. 
  \sqrt{\frac{L+1}{2L+3}}
 \left[\begin{array}{ccc}
       L & S & J \\
       1 & 1 & 0 \\
     L+1 & S' & J
        \end{array}\right] {\mathcal Y}_{J L+1 S'}^{M}(\hat{\bf p}) \right\}
\left\{\delta_{S',1}\delta_{S,0}-\sqrt{3}\delta_{S',0}\delta_{S,1}\right\}\ .
\nonumber \\ 
\label{eq:appb.92} \end{eqnarray}
Working this out explicitly, we find
\begin{eqnarray}
\left({\bf AS}\cdot\hat{\bf p}\right) {\mathcal Y}_{J J-1 1}^{M}(\hat{\bf p})&=&
    i\ b_{J}\ {\mathcal Y}_{J J 0}^{M}(\hat{\bf p}) \ ,\nonumber \\
\left({\bf AS}\cdot\hat{\bf p}\right) {\mathcal Y}_{J J+1 1}^{M}(\hat{\bf p})&=&
  \hspace*{0.3cm} i\ a_{J}\ {\mathcal Y}_{J J 0}^{M}(\hat{\bf p}) \ ,\nonumber\\
\left({\bf AS}\cdot\hat{\bf p}\right) {\mathcal Y}_{J\ J\ 1}^{M}(\hat{\bf p})&=&0 \ ,\nonumber \\
\left({\bf AS}\cdot\hat{\bf p}\right) {\mathcal Y}_{J\ J\ 0}^{M}(\hat{\bf p})&=&
  \hspace*{0.3cm}  -i\ b_{J}\ {\mathcal Y}_{J J-1 1}^{M}(\hat{\bf p}) -i\
    a_{J}\ {\mathcal Y}_{J J+1 1}^{M}(\hat{\bf p})\ .           
\label{eq:appb.93} \end{eqnarray}
{}From the results above one can derive the following useful
partial wave projections. For the spin triplet states:
\begin{eqnarray}
 ( L' 1 J| V({\bf k}^{2})\left({\bf S}\cdot\hat{\bf p}_i\right)^{2}
      | L 1 J )&=& 4\pi
  \left(\begin{array}{ccc}
  a_{J}^{2}V_{J-1} & 0 & - a_{J}b_{J}V_{J-1} \\[0.2cm]
   0 & V_{J}& 0 \\[0.2cm]
  - a_{J}b_{J}V_{J+1} & 0 & b_{J}^{2}V_{J+1}   \end{array}
  \right) \ , \nonumber
\label{eq:appb.7a} \\
 ( L' 1 J|({\bf S}\cdot\hat{\bf p}_f)^{2} V({\bf k}^{2})
      | L 1 J )&=& 4\pi
  \left(\begin{array}{ccc}
  a_{J}^{2}V_{J-1} & 0 & - a_{J}b_{J}V_{J+1} \\[0.2cm]
   0 & V_{J}& 0 \\[0.2cm]
  - a_{J}b_{J}V_{J-1} & 0 & b_{J}^{2}V_{J+1}   \end{array}
  \right) \ , \nonumber \label{eq:appb.7b} \\
 ( L' 1 J|({\bf S}\cdot\hat{\bf p}_f) V({\bf k}^{2})
 ({\bf S}\cdot\hat{\bf p}_i) | L 1 J ) &=&
4\pi
  \left(\begin{array}{ccc}
  a_{J}^{2}V_{J} & 0 & - a_{J}b_{J}V_{J} \\[0.2cm]
  0 &  a_{J}^{2}V_{J-1}+b_{J}^{2}V_{J+1}& 0 \\[0.2cm]
  - a_{J}b_{J}V_{J} & 0 & b_{J}^{2}V_{J}   \end{array}
  \right)   \ , \nonumber
\\
 ( L' 1 J| V({\bf k}^{2})\left({\bf AS}\cdot\hat{\bf p}_i\right)^{2}
      | L 1 J )&=& 4\pi
  \left(\begin{array}{ccc}
  b_{J}^{2}V_{J-1} & 0 &  a_{J}b_{J}V_{J-1} \\[0.2cm]
   0 & 0 & 0 \\[0.2cm]
   a_{J}b_{J}V_{J+1} & 0 & a_{J}^{2}V_{J+1}   \end{array}
  \right) \ , \nonumber
\label{eq:appb.7aa} \\
 ( L' 1 J|({\bf AS}\cdot\hat{\bf p}_f)^{2} V({\bf k}^{2})
      | L 1 J )&=& 4\pi
  \left(\begin{array}{ccc}
  b_{J}^{2}V_{J-1} & 0 & a_{J}b_{J}V_{J+1} \\[0.2cm]
   0 & 0 & 0 \\[0.2cm]
  a_{J}b_{J}V_{J-1} & 0 & a_{J}^{2}V_{J+1}   \end{array}
  \right) \ , \nonumber \label{eq:appb.7bb} \\
 ( L' 1 J|({\bf AS}\cdot\hat{\bf p}_f) V({\bf k}^{2})
 ({\bf AS}\cdot\hat{\bf p}_i) | L 1 J ) &=&
4\pi
  \left(\begin{array}{ccc}
  b_{J}^{2}V_{J} & 0 & a_{J}b_{J}V_{J} \\[0.2cm]
  0 & 0 & 0 \\[0.2cm]
   a_{J}b_{J}V_{J} & 0 & a_{J}^{2}V_{J}   \end{array}
  \right)\ .        
\label{eq:appb.7c} 
\end{eqnarray}
For the spin singlet states:
\begin{eqnarray}
 ( J 0 J| V({\bf k}^{2})\left({\bf AS}\cdot\hat{\bf p}_i\right)^{2}
      | J 0 J )&=& 4\pi\ V_{J}\ , \nonumber
\label{eq:appb.7aaa} \\
 ( J 0 J|({\bf AS}\cdot\hat{\bf p}_f)^{2} V({\bf k}^{2})
      | J 0 J )&=& 4\pi \ V_{J}\ , \nonumber
\label{eq:appb.7bbb} \\
 ( J 0 J|({\bf AS}\cdot\hat{\bf p}_f) V({\bf k}^{2})
 ({\bf AS}\cdot\hat{\bf p}_i) | J 0 J ) &=& 4\pi \left(b_{J}^{2}V_{J-1} + a_{J}^2V_{J+1}\right)\ .       
\label{eq:appb.7cc} 
\end{eqnarray}
For the spin singlet-triplet transitions:
\begin{eqnarray}
 ( J 1 J|({\bf S}\cdot\hat{\bf p}_f) V({\bf k}^{2})
 ({\bf AS}\cdot\hat{\bf p}_i) | J 0 J ) &=& -4\pi \ a_Jb_J\left(V_{J-1} - V_{J+1}\right)\ , \nonumber
\label{eq:appb.7ccc} \\
 ( J 0 J|({\bf AS}\cdot\hat{\bf p}_f) V({\bf k}^{2})
 ({\bf S}\cdot\hat{\bf p}_i) | J 1 J ) &=& -4\pi \ a_Jb_J\left(V_{J-1} - V_{J+1}\right)\ .    
\label{eq:appb.7cccc} 
\end{eqnarray}
Using the identity
\begin{equation}
(\mbox{\boldmath $\sigma$}_1\cdot{\bf a})(\mbox{\boldmath $\sigma$}_2\cdot{\bf a})=
      2({\bf S}\cdot{\bf a})^{2} - {\bf a}^{2} \ ,
\label{eq:appb.9} \end{equation}
the spinor invariants $P_2$-- $P_8$ can be written as
\begin{eqnarray}
P_2&=&2{\bf S}^2-3 \ ,\nonumber \\
P_3&=&2\left[\left({\bf S}\cdot{\bf p}_f\right)^2+\left({\bf S}\cdot{\bf p}_i\right)^2-\left({\bf S}\cdot{\bf p}_f\right)\left({\bf S}\cdot{\bf p}_i\right)+\left({\bf AS}\cdot{\bf p}_f\right)\left({\bf AS}\cdot{\bf p}_i\right)-{\bf p}_f\cdot{\bf p}_i\right] \nonumber \\
&& -\frac{2}{3}{\bf S}^2\left(p_f^2+p_i^2-2{\bf p}_f\cdot{\bf p}_i\right) \ ,\nonumber \\
P_4&=&-\left[\left({\bf S}\cdot{\bf p}_f\right)\left({\bf S}\cdot{\bf p}_i\right)+\left({\bf AS}\cdot{\bf p}_f\right)\left({\bf AS}\cdot{\bf p}_i\right)-{\bf p}_f\cdot{\bf p}_i\right] \ ,\nonumber \\
P_{5} &=&\left(2 {\bf S}^{2} - 1 \right) \left({\bf p}_f\times{\bf p}_i\right)^{2}-2p_f^{2} p_i^{2}\left[\left({\bf S}\cdot\hat{\bf p}_f\right)^{2}+\left({\bf S}\cdot\hat{\bf p}_i\right)^{2}\right]
   \nonumber \\[0.2cm]  & & +
 2\left[\left({\bf S}\cdot{\bf p}_f\right)\left({\bf S}\cdot{\bf p}_i\right)
 -\left({\bf AS}\cdot{\bf p}_f\right)\left({\bf AS}\cdot{\bf p}_i\right) + {\bf p}_f\cdot{\bf p}_i\right]
 \left({\bf p}_f\cdot{\bf p}_i\right)
 \ ,\nonumber \\
P_6&=&-\left[\left({\bf S}\cdot{\bf p}_f\right)\left({\bf AS}\cdot{\bf p}_i\right)+\left({\bf AS}\cdot{\bf p}_f\right)\left({\bf S}\cdot{\bf p}_i\right)\right] \ ,\nonumber \\
P_7&=&\left[2\left({\bf S}\cdot{\bf p}_f\right)^2-2\left({\bf S}\cdot{\bf p}_i\right)^2-{\bf p}_f^2+{\bf p}_i^2\right] \ ,\nonumber \\
P_8&=&\frac{1}{2}\left[\left({\bf S}\cdot{\bf p}_f\right)\left({\bf AS}\cdot{\bf p}_i\right)-\left({\bf AS}\cdot{\bf p}_f\right)\left({\bf S}\cdot{\bf p}_i\right)\right]\ .
\end{eqnarray}
For $P_5$ we use $\left({\bf p}_f\times{\bf p}_i\right)^2=q_f^2q_i^2\left(1-x^2\right)$, where $x=\hat{{\bf p}}_f\cdot\hat{{\bf p}}_i$. In case of an extra
factor $\left({\bf p}_f\cdot{\bf p}_i\right)$, as occurs for example
in the second line of $P_5$, we simply use
the expansion
\begin{equation}
 \left({\bf p}_f\cdot{\bf p}_i\right) V({\bf k}^{2})= p_f p_i
 \sum_{L=0}^{\infty}(2L+1)\tilde{V}_{L}(x) P_{L}(\cos \theta) \ ,
\label{eq:appb.14} \end{equation}
where
\begin{equation}
 \tilde{V}_{L}=\frac{1}{2L+1}\left[(L+1)V_{L+1}+L V_{L-1}\right]\ .
\label{eq:appb.15} \end{equation}

\bibliographystyle{elsart-num}

\bibliography{yneft}

\end{document}